\pdfoutput=1
\RequirePackage{ifpdf}
\ifpdf 
\documentclass[pdftex]{sigma}
\else
\documentclass{sigma}
\fi

\numberwithin{equation}{section}
\DeclareMathOperator{\Sp}{Sp}
\DeclareMathOperator{\D}{d}

\begin{document}

\allowdisplaybreaks

\renewcommand{\PaperNumber}{066}

\FirstPageHeading

\ShortArticleName{Symmetry and Intertwining Operators}

\ArticleName{Symmetry and Intertwining Operators\\
for the Nonlocal Gross--Pitaevskii Equation}

\Author{Aleksandr L.~LISOK~$^\dag$, Aleksandr V.~SHAPOVALOV~$^{\dag\ddag}$ and Andrey Yu.~TRIFONOV~$^{\dag\ddag}$}

\AuthorNameForHeading{A.L.~Lisok, A.V.~Shapovalov and A.Yu.~Trifonov}

\Address{$^\dag$~Mathematical Physics Department, Tomsk Polytechnic University,
\\
\hphantom{$^\dag$}~30 Lenin Ave., Tomsk, 634034 Russia}
\EmailD{\href{mailto:lisok@tpu.ru}{lisok@tpu.ru}, \href{mailto:atrifonov@tpu.ru}{atrifonov@tpu.ru}}

\Address{$^\ddag$~Theoretical Physics Department, Tomsk State University,
\\
\hphantom{$^\ddag$}~36 Lenin Ave., Tomsk, 634050 Russia}
\EmailD{\href{mailto:shpv@phys.tsu.ru}{shpv@phys.tsu.ru}}

\ArticleDates{Received February 15, 2013, in f\/inal form October 26, 2013; Published online November 06, 2013}

\Abstract{We consider the symmetry properties of an integro-dif\/ferential multidimensional
Gross--Pitaevskii equation with a~nonlocal nonlinear (cubic) term in the context of sym\-metry analysis
using the formalism of semiclassical asymptotics.
This yields a~semiclassically reduced nonlocal Gross--Pitaevskii equation, which can be treated as a~nearly
linear equation, to determine the principal term of the semiclassical asymptotic solution.
Our main result is an approach which allows one to construct a~class of symmetry operators for the reduced
Gross--Pitaevskii equation.
These symmetry operators are determined by linear relations including intertwining operators and additional
algebraic conditions.
The basic ideas are illustrated with a~1D reduced Gross--Pitaevskii equation.
The symmetry operators are found explicitly, and the corresponding families of exact solutions
are obtained.}


\Classification{35Q55; 45K05; 76M60; 81Q20}

\def\Im{\mathop{\rm Im}\nolimits} \def\Re{\mathop{\rm Re}\nolimits}

\section{Introduction}

\looseness=1
Symmetry operators, which, by def\/inition, leave the set of solutions of an equation invariant, are of
essential importance in the symmetry analysis of nonlinear partial dif\/ferential equations (PDEs).
The obvious use of symmetry operators of an equation is to generate new solutions from a~known one.
A modern symmetry analysis of dif\/ferential equations (DEs) is based on Lie group theory.
For example, if for an ordinary dif\/ferential equation (ODE) there exists a~Lie group of point
transformations (point symmetries) which act on the space of independent and dependent variables, then they map
any solution to another solution of the equation.
In a~more general case, an ODE can admit contact transformations (contact symmetries) acting on the
independent and dependent variables, and  also  on the f\/irst derivatives of the dependent variables.
In other words, point symmetries and contact symmetries provide examples of Lie groups of symmetry
operators.
The Lie group methods,
as well as their applications to ODEs and PDEs,
are described in many books and review articles
(see, e.g.,~\cite{BlumAnco}).
The prolongation of the action of a~Lie group on the space of independent variables, dependent variables, and
partial derivatives of the dependent variables up to any f\/inite order allows to apply the Lie group
theory to
studying symmetries of PDEs~\cite{shapovalov:OVS}.
The fundamental property of a~Lie group is that it is completely characterized by its inf\/inite\-si\-mal
operator (generator).
Given a system of DEs, f\/inding the Lie symmetry group is reduced to solving a~system of
equations that determine
the Lie group generators.
The principal point is that the determining equations for the generator are linear and homogeneous.
For a~nonlinear PDE the determining equations take the form of an overdetermined system of linear
homogeneous PDEs, which can be solved step-by-step to obtain
inf\/initesimal operators in explicit form
(see, e.g.,~\cite{BlumChevAnco,shapovalov:OLVER, shapovalov:OVS}).

\looseness=1
Solving the determining equations for given system of PDEs
we can f\/ind the generators of point or
contact symmetries for the system.
Following the Lie theory, we can recover the Lie group of f\/inite (i.e., not inf\/initesimal) symmetry
transformations for given system of PDEs.
Solutions of the determining equations, however, may contain not only independent variables, dependent
variables, and f\/irst order derivatives (as with point and contact symmetries), but also higher-order
derivatives.
Generators of this type are called higher symmetries and they do not yield f\/inite Lie groups.
Higher symmetries are related to the so-called Lie--B\"acklund transformations that are widely used in
symmetry analysis (see~\cite{shapovalov:IBRAGIM} and also, e.g.,~\cite{shapovalov:OLVER}
and~\cite{BlumChevAnco}).
Note that higher-order symmetries do not generate symmetry operators.
However, no general approaches to
direct calculation of
symmetry operators for nonlinear equations are
known other than the use of the Lie group formalism.
This is due to that the determining equations for symmetry operators are nonlinear operator equations.
Solving them is a~complicated mathematical problem which requires special techniques not developed yet.
In addition, in order to solve
determining equations for symmetry operators, we have to specify the structure of
symmetry operators consistent
with
the determining equations, but there are no recipes for choosing such a~structure.
Therefore, f\/inding the symmetry operators for nonlinear equations is in general
an unrealistic task.

\looseness=1
Note that for linear PDEs, symmetry operators which are widely used in quantum mechanics applications
can be ef\/fectively found from linear determining equations,
(see, e.g.,~\cite{shapovalov:FUSCH-N,shapovalov:MANKO,Perel} and references therein).
This inspired us
to seek a~special class of nonlinear equations for which symmetry operators could be
calculated using the methods applicable to linear equations.
As an example of such a~class of nonlinear equations
we consider nonlinear integro-dif\/ferential
equations (IDEs) with partial derivatives.
We call the equations of this class
\emph{nearly linear equations}.
Symmetry operators for them
can be found by solving linear operator equations (similarly to those
for linear PDEs) and additional algebraic equations.
We consider a~generalized multidimensional integro-dif\/ferential Gross--Pitaevskii equation (GPE) with
partial derivatives and a~nonlocal cubic nonlinear interaction term of general form.
The WKB--Maslov method of semiclassical asymptotics~\cite{shapovalov:BEL-DOB, shapovalov:MASLOV-1} is used
to obtain a~reduced GPE from the original GPE.
The reduced GPE is quadratic in spatial coordinates and derivatives, and it contains a~nonlocal cubic
nonlinear interaction term of special form.
This equation belongs to the class of nearly linear equations and determines the principal term
of semiclassical asymptotic solution.

The main result of our work is an approach developed for f\/inding symmetry operators for a~reduced GPE by
solving linear operator equations.
This approach is illustrated by
an example of a~one-dimensional reduced GPE
for which
symmetry operators can be found explicitly.
Using symmetry operators obtained
two families of exact solutions can be generated for the reduced GPE.
In Section~\ref{section2} the
integro-dif\/ferential Gross--Pitaevskii equation is considered and its
semiclassical reduction is presented.
A method for integrating the reduced GPE is described and the essential idea of the method is realized;
namely, the consistent system and the linear equation associated with the reduced GPE are found.
In Section~\ref{section3} we propose an approach
to
f\/inding the class of symmetry operators of the reduced GPE by
constructing intertwining operators.
The general ideas are illustrated in Section~\ref{section4} by the example of a~one-dimensional GPE of
special type.
The symmetry operators for this equation are found explicitly, and two families of exact solutions are generated
making use of the operators obtained.

\newpage

\section{The nonlocal Gross--Pitaevskii equation\\and the Cauchy problem}\label{section2}

We consider here
the Gross--Pitaevskii equation with a~nonlocal interaction term of general form.
Using the concepts of the semiclassical WKB--Maslov method, we arrive at a~reduced nonlocal GPE and
brief\/ly explain
an algorithm for solving the Cauchy problem.

The Gross--Pitaevskii equation and its modif\/ications are widely used in study of coherent matter waves in
Bose--Einstein condensates (BECs)~\cite{lisok:CORNELL}.
Recent extensions to BEC studies involve long-range ef\/fects in the condensates described by a~generalized
GPE containing integral terms responsible for nonlocal interactions.
We refer to equations of this class as nonlocal GPE (which are also known as Hartree-type equations).
The nonlocal BEC models may keep the condensate wave function from collapse and stabilize the solutions in
higher dimensions (see, e.g.,~\cite{shapovalov:LUSHNIKOV}, the review~\cite{shapovalov:FRANTZESKAKIS} and
references therein).
Nonlocal GPEs also serve as basic equations
of models describing many-particle quantum systems, nonlinear
optics phenomena~\cite{AGRAWAL}, collective soliton excitations in atomic chains~\cite {NOVOA}, etc.

Let us write
the nonlocal Gross--Pitaevskii equation as
\begin{gather}
\hat F(\Psi)(\vec x,t)=\{{-}i\hbar\partial_t+\hat H(t)+\varkappa\hat V(\Psi)(t)\}\Psi(\vec x,t)=0,
\label{shapovalov:GPE}
\\
\hat V(\Psi)(t)=V(\Psi)(\hat z,t)=\int_{{\mathbb R}^n}
\D\vec y\, \Psi^*(\vec y,t)V(\hat z,\hat w,t)\Psi(\vec y,t),
\label{shapovalov:VTE-2}
\end{gather}
where $\partial_t=\partial/\partial t$, $\Psi(\vec x,t)$ is a~smooth complex scalar function that
belongs to a~complex Schwartz space $\mathbb{S}$ in the space variable $\vec x \in {\mathbb R}^n$ at each time $t$.

The linear operators $\hat H(t)= H (\hat z,t)$ and $V(\hat z,\hat w,t)$ in~\eqref{shapovalov:GPE} are
Hermitian Weyl-ordered functions~\cite{shapovalov:KARASEVMASLOV} of time $t$ and of noncommuting operators
\begin{gather*}
\hat z=(\hat{\vec p},\vec x)=(-i\hbar\partial/{\partial\vec x},\vec x),
\qquad
\hat w=(-i\hbar\partial/{\partial\vec y},\vec y),
\qquad
\vec x,\vec y\in{\mathbb R}^n,
\end{gather*}
with the commutators
\begin{gather*}
[\hat z_k,\hat z_j]_-=[\hat w_k,\hat w_j]_-=i\hbar J_{kj},
\qquad
[\hat z_k,\hat w_j]_-=0,
\qquad
k,j=\overline{1,2n},
\end{gather*}
where $[\hat A,\hat B]_{-}=\hat A\hat B-\hat B\hat A$, $J =\|J_{kj}\|_{2n\times 2n}$ is
the~unit symplectic matrix: $J=
\begin{pmatrix}
0&-{\mathbb I}
\nonumber\\
{\mathbb I}& 0
\end{pmatrix}
_{2n\times 2n}$, and ${\mathbb I}={\mathbb I}_{n\times n}$ is the
$n\times n$ identity matrix.
We use the space $\mathbb{S}$ to provide
existence of the moments of $\Psi(\vec x,t)$ and convergence
of the integral in~\eqref{shapovalov:VTE-2}.
In what follows, we use the norm~$\|\Psi\|$, $\Psi \in \mathbb{S}$, of the space $L_2({\mathbb R}^n_x)$,
i.e., $\|\Psi\|=\sqrt{(\Psi,\Psi)}$, where 
$(\Phi,\Psi)=\displaystyle\int_{{\mathbb R}^n} \D\vec x
\Phi^*(\vec x)\Psi(\vec x)$ denotes the Hermitian inner product of the functions $\Phi,\Psi\in \mathbb{S}$,
and $\Phi^*$ denotes the complex conjugate to~$\Phi$.

From equation~\eqref{shapovalov:GPE} it follows immediately that the squared norm of a~solution $\Psi(\vec x,t)$ is
conserved, $\|\Psi (t)\|^2=\|\Psi (0)\|^2=\rm{const}$.

A specif\/ic and attractive feature of the nonlocal GPE~\eqref{shapovalov:GPE} is that 
in the semiclassical appro\-xi\-ma\-tion
the input GPE is reduced to an equation containing nonlocal terms which
can be expressed as a~f\/inite number of moments of the unknown function $\Psi(\vec x,t)$.
The reduced equation can be considered as
nearly linear.
The concept of
the nearly linear equations
implies that among the solutions of a~nonlinear equation
there exists a~subset of solutions that regularly depend
on the nonlinearity parameter~\cite{shapovalov:KARASEVMASLOV}.
In the multidimensional case, the GPE~\eqref{shapovalov:GPE} with variable coef\/f\/icients of general form
cannot be integrated by well-known methods, such as the inverse scattering
transform~\cite{shapovalov:ZAKHAROV-2}.
Therefore, analytical solutions to this equation can be constructed only approximately.
An ef\/fective approach to constructing asymptotic solutions in this case is
to f\/ind semiclassical asymptotics as $\hbar\to 0$.

Note that semiclassical asymptotic expansions can be assigned to the following basic classes.
The semiclassical asymptotic solutions of the equation under consideration are constructed in a~chosen
class of functions $K_\hbar$.
The functions of the class $K_\hbar$ are determined by specif\/ic features of the problem and singularly
depend on the small parameter $\hbar$.
In the general case, such a~class of functions is constructed as follows: In the phase space of a~dynamic
system of equations corresponding to the equation with partial derivatives under consideration (the
classical equations of motion in the case of a~linear quantum mechanics Schr\"odinger equation),
a~Lagrangian manifold $\Lambda^k$, $k\le n$, is def\/ined.
Here
$k$ is the dimension of $\Lambda^k$ and $n$ is the dimension of the conf\/iguration space of the
phase space.
The manifold $\Lambda^k$ evolves in time for the Cauchy problem and is invariant for the spectral problem,
i.e.\
 $\Lambda^k$ is not deformed and does not move in space.
On the manifold $\Lambda^k$ 
a~set of functions is def\/ined.
The Maslov's canonical operator projects a~function def\/ined in
the phase space onto a~function given in
the conf\/iguration space.
If $k=n$, then the canonical operator should be a~real phase operator~\cite{MasFed}, whereas if $k<n$, then
the canonical operator should be a~complex phase one~\cite{shapovalov:BEL-DOB, shapovalov:MASLOV-1}.
In constructing projections of $\Lambda^k$ onto
the conf\/iguration space,
caustics can appear.

The solutions of the f\/irst class ($k = n$) are given by the WKB ansatz with a~real phase~\cite{MasFed},
where the leading term of the asymptotics outside the neighborhoods of the focal points can be written as
\begin{gather}
\Psi(\vec x,t,\hbar)=\sum_{j=1}^Mf_j(\vec x,t)\exp\Big\{\frac i\hbar S_j(\vec x,t)\Big\}e^{i\pi\mu_j/2},
\label{shapovalov:GPE8-1aaa}
\\
\mu_j\in\mathbb{Z},
\qquad
\Im S_j(\vec x,t)=0,
\qquad
f_j,S_j\in\mathbb{C}^\infty(\mathbb{R}^n).
\nonumber
\end{gather}

Semiclassical asymptotic solutions of the form~\eqref{shapovalov:GPE8-1aaa} for the Gross--Pitaevskii
equation were const\-ructed in~\cite{Mas3, Mas1,Mas2} (see also~\cite{Karas}).

The solutions of the second class ($k=0$) are constructed using a~complex WKB--Maslov
ansatz~\cite{shapovalov:BEL-DOB, shapovalov:MASLOV-1}.
For the Gross--Pitaevskii equation, asymptotic solutions of this type are considered
in~\cite{shapovalov:BTS1, DobrShafarNekr, Vacul}.

Note that constructing
of semiclassical asymptotic solutions for nonlinear equations
engender
a~number of problems: In general, the evolution law for a~manifold $\Lambda^k$ is unknown.
In other words, the ``classical dynamics'' related to the nonlinear equation under consideration depends on
the initial conditions for the equation.
Moreover, the relevant ``classical dynamics equations'' are unknown a~priori for the nonlinear equation and
to deduce them is a~real problem.
For the Gross--Pitaevskii equation~\eqref{shapovalov:GPE}, this problem was solved for the class of
functions concentrated on a~zero-dimensional manifold $\Lambda^0$~\cite{shapovalov:BTS1} and for the class
of functions concentrated on an $n$-dimensional manifold $\Lambda ^n$~\cite{Mas3,Mas1,Mas2}.

Following~\cite{shapovalov:BTS1}, we denote the second class of functions by ${\mathcal
P}_{\hbar}^t(Z(t,{\hbar}), S(t,{\hbar}))$ and def\/ine it as
\begin{gather*}
{\mathcal P}_{\hbar}^t
={\mathcal P}_{\hbar}^t\big(Z(t,{\hbar}),S(t,{\hbar})\big)
=\biggl\{\!\Phi:\Phi(\vec x,t,{\hbar})\!
=\!{\varphi}\left(\!\frac{{\Delta}\vec x}{\sqrt{\hbar}}
,t,{\hbar}\!\right)
\exp\Bigl[\displaystyle\frac{i}{\hbar}(S(t,{\hbar})+{\langle}\vec P(t,{\hbar}),{\Delta}
\vec x{\rangle})\Bigr]\!\biggr\},
\end{gather*}
where the function ${\varphi}(\vec\xi,t,{\hbar})$ belongs to the Schwarz space $\mathbb S$ in the variable
$\vec\xi\in{\mathbb R}^n $, smoothly depends on $t$, and regularly depends on $\sqrt{\hbar}$ as
${\hbar}\to0$.
Here $\Delta\vec x=\vec x-\vec X(t,{\hbar})$, and the real function $S(t,{\hbar})$ and the $2n$-dimensional
vector function $Z(t,{\hbar})=(\vec P(t,{\hbar}), \vec X(t,{\hbar}))$, which characterize the class
${\mathcal P}_{\hbar}^t(Z (t,{\hbar}), S(t,{\hbar}))$, regularly depend on $\sqrt{\hbar}$ in the
neighborhood of ${\hbar}=0$ and {\em are to be determined}.
Note that ${\mathcal P}_{\hbar}^t(Z (t,{\hbar}), S(t,{\hbar})) \subset \mathbb{S}$.
If this does not lead to misunderstanding, we use the contracted notation ${\mathcal P}_{\hbar}^t$ for
${\mathcal P}_{\hbar}^t(Z(t,{\hbar}),S(t,{\hbar}))$.

Here to construct symmetry operators
we use the complex WKB--Maslov asymptotic solutions of the Cauchy
problem for the GPE~\eqref{shapovalov:GPE}
\begin{gather}
\Psi(\vec x,t)\big|_{t=s}=\psi(\vec x),
\qquad
\psi(\vec x)\in{\mathcal P}_{\hbar}^0,
\label{cauchy-GPE}
\end{gather}
where
\begin{gather*}
{\mathcal P}_{\hbar}^0={\mathcal P}_{\hbar}^0\big(Z_0({\hbar}),S_0({\hbar}
)\big)=\biggl\{\phi:\phi(\vec x,{\hbar})={\varphi}\left(\frac{{\Delta}\vec x}{\sqrt{\hbar}}
,t,{\hbar}\right)\exp\Bigl[\displaystyle\frac{i}{\hbar}(S_0({\hbar})+{\langle}\vec P_0({\hbar}
),{\Delta}\vec x_0{\rangle})\Bigr]\biggr\},
\\
Z_0({\hbar})=\big({\vec P}_0({\hbar}),{\vec X}_0({\hbar})\big),
\qquad
{\Delta}\vec x_0=\vec x-{\vec X}_0({\hbar}).
\end{gather*}

The def\/inition of the class of trajectory-concentrated functions contains the phase trajectory
$Z(t,{\hbar})$ and the scalar function $S(t,{\hbar})$ as ``free parameters''.
The functions belonging to the class~${\mathcal P}_{\hbar}^t$, at any f\/ixed time $t\in{\mathbb R}^1$ are
{\em concentrated}, as ${\hbar}\to0$ in the neighborhood of a~point lying on the phase curve $z=Z(t,0)$,
$t\in{\mathbb R}^1$~\cite{Bagre}.
Therefore, it is natural to call the functions of the class~${\mathcal P}_{\hbar}^t$
{\em trajectory-concentrated functions}.

The WKB solutions of the form~\eqref{shapovalov:GPE8-1aaa} are concentrated on a~family of phase
trajectories whose projections on the conf\/iguration space may intersect, giving rise to a~caustic
problem~\cite{MasFed}.
On the other hand, all semiclassical asymptotics of the class ${\mathcal P}_{\hbar}^t$ are concentrated on
the same trajectory.
So we do not face problems with caustics and collapse problem in constructing trajectory-concentrated
solutions of the GPE.

Let $\widehat O(\hbar^\nu)$
be
an operator $\hat F$ such that for any function $\Phi$ belonging to
the space ${\mathcal P}_{\hbar}^t$ the following asymptotic estimate is valid:
\begin{gather*}
\frac{\|\hat F\Phi\|}{\|\Phi\|}=O(\hbar^\nu),
\qquad
\hbar\to0,
\qquad
\Delta\hat z=\hat z-Z(t,\hbar)
\end{gather*}

It may be shown (see~\cite{Bagre,shapovalov:BTS1}) that for the functions belonging to ${\mathcal
P}_{\hbar}^t$, the following asymptotic estimate
is valid:
\begin{gather}
\Delta{\hat z}=\widehat O\big({\hbar}^{1/2}\big),
\qquad
\hbar\to0.
\label{bbst1.9}
\end{gather}

Let us expand the operators $\hat H(t)= H (\hat z,t)$ and $\hat V(t)=V(\hat z,\hat w,t)$
in~\eqref{shapovalov:GPE} as Taylor series in the operators
$\Delta\hat z=\hat z-Z(t,\hbar)$ and
$\Delta\hat w=\hat w-Z(t,\hbar)$, respectively, and restrict ourselves to quadratic terms.
Then, in view of~\eqref{bbst1.9}, the solution of the Cauchy problem~\eqref{shapovalov:GPE}
and~\eqref{cauchy-GPE} asymptotic in a~formal small parameter $\hbar$ ($\hbar\to0$)
can be constructed\footnote{Note that in the semiclassical trajectory-coherent approximation, if $\hbar$ is small
enough ($\hbar\to 0$), all results are established for a~f\/inite time interval $[0, T]$.
Evidently, this version of the semiclassical approach is not uniform in time as $T\to\infty$
(see~\cite{Bagre,shapovalov:BTS1}).
Therefore, the problem of long-time validity of the semiclassical trajectory coherent approximation (i.e.,
the two limits, as $\hbar\to0$ and then as $T\to\infty$) should be the subject of special study.} accurate
to $O(\hbar^{3/2})$ (see~\cite{shapovalov:BTS1}).
The leading-order term of the asymptotics
can be found by reducing the GPE~\eqref{shapovalov:GPE} to a~GPE
with a~quadratic nonlocal operator.

The higher-order corrections to the leading-order term can be found using perturbation theo\-ry~\cite{shapovalov:BTS1}.
Thus the study of GPEs with a~quadratic nonlocal operator is crucial for the construction of semiclassical
asymptotics for this type of GPE in the class of trajectory concentrated functions.
Without loss of generality, we consider a~GPE of the form
\begin{gather}
\left\{{-}i\hbar\partial_t+\hat H_{\rm qu}(\hat z,t)+\varkappa\int_{{\mathbb R}^n}
\D\vec y\Psi^*(\vec y,t)V_{\rm qu}(\hat z,\hat w,t)\Psi(\vec y,t)\right\}\Psi(\vec x,t)=0,
\label{QUAD-HAMILT}
\end{gather}
where the linear operators $H_{\rm qu}(\hat z,t)$ and $V_{\rm qu}(\hat z,\hat w,t)$ are Hermitian and
quadratic in $\hat z$, $\hat w$, respectively:
\begin{gather}
H_{\rm qu}(\hat z,t)=\dfrac{1}{2}\langle\hat z,{\mathcal H}_{zz}(t)\hat z\rangle+\langle{\mathcal H}
_z(t),\hat z\rangle,
\label{shapovalov:QUAD-1}
\\
V_{\rm qu}(\hat z,\hat w,t)=\dfrac{1}{2}\langle\hat z,W_{zz}(t)\hat z\rangle+\langle\hat z,W_{zw}
(t)\hat w\rangle+\dfrac{1}{2}\langle\hat w,W_{ww}(t)\hat w\rangle.
\label{shapovalov:QUAD-2}
\end{gather}
Here ${\mathcal H}_{zz}(t)$, $W_{zz}(t)$, $W_{zw}(t)$, and $W_{ww}(t)$ are $2n\times 2n$ matrices;
${\mathcal H}_z(t)$ is a~$2n$ vector; the angle brackets $\langle\cdot,\cdot\rangle$ denote the Euclidean inner
product of vectors:
\begin{gather*}
\langle\vec p,\vec x\rangle=\sum\limits^n_{j=1}p_jx_j,
\quad
\vec p,\vec x\in {\mathbb R}^n;
\qquad
\langle z,w\rangle=\sum\limits^{2n}_{j=1}z_jw_j,
\quad
z,w\in {\mathbb R}^{2n}.
\end{gather*}
We call equation~\eqref{QUAD-HAMILT} with the linear operators $H_{\rm qu} $ and $V_{\rm qu}$ given
by~\eqref{shapovalov:QUAD-1} and~\eqref{shapovalov:QUAD-2}, respectively, {\it a~reduced Gross--Pitaevskii
equation} (RGPE).

An RGPE can be integrated explicitly~\cite{shapovalov:LTS, Lisok:Sigma} and it possesses very rich symmetries.
Analysis of these symmetries can provide a~wealth of information about the equation and its solutions.

As an RGPE contains a~nonlocal nonlinear term, its symmetry properties are of special interest
in the
symmetry analysis of partial dif\/ferential equations.
The matter is that the application of the standard methods of symmetry analysis~\cite{shapovalov:IBRAGIM, shapovalov:FUSCH-N, shapovalov:OLVER,shapovalov:OVS}, developed basically for PDEs,
leads to a~number of dif\/f\/iculties when applied to equations dif\/ferent from PDEs:
For instance, there are no regular rules for choosing an appropriate structure of symmetries for non-dif\/ferential equations.
This problem can be avoided by using an RGPE as its symmetry properties are closely related to the symmetry
of the linear equation associated with the input nonlinear equation.

The key factor in symmetry analysis
of the nonlinear equation $\hat F(\Psi)(\vec x,t)=0$ is the
symmetry operator $\hat A$ that makes the set of solutions of the equation invariant (see, e.g.,~\cite{shapovalov:MANKO}):
\begin{gather}
\hat F(\Psi)(\vec x,t)=0\quad\Rightarrow\quad\hat F(\hat A\Psi)(\vec x,t)=0.
\label{symm-oper-1}
\end{gather}
Generally, it is impossible to f\/ind ef\/fectively
a~symmetry operator $\hat A$ for a~given nonlinear
operator $\hat F$ by solving the nonlinear operator equation~\eqref{symm-oper-1}.
This situation is resolved in the group analysis of dif\/ferential equations~\cite{shapovalov:IBRAGIM, shapovalov:OLVER, shapovalov:OVS} where a~symmetry $\hat\sigma$
(generator of a~Lie group of symmetry operators) is the main object of analysis.

The symmetries are determined by the linear operator equation
\begin{gather*}
\hat F(\Psi)(\vec x,t)=0\quad\Rightarrow\quad\hat F'(\hat\sigma\Psi)(\vec x,t)=0.
\end{gather*}
Here $\hat F'(\Psi)$ is the Freshet derivative of $\hat F$ calculated for $\Psi$.
For a~linear operator $\hat F$, we have $\hat F'=\hat F$ and the symmetry operators being the same as the symmetries.

We assign the RGPE~\eqref{QUAD-HAMILT} to the class of nearly linear equations,
following the def\/inition given in~\cite{LevShTr12}:
A nearly linear equation determining a~function $\Psi$ has the form of
a~linear partial dif\/ferential equation with coef\/f\/icients depending on the moments of the function
$\Psi$.
This type of equation can be associated with a~consistent system which includes a~system of ordinary
dif\/ferential equations (ODEs) describing the evolution of the moments and RGPE.

Using the RGPE as an example, we can see that the class of symmetry operators for nearly linear equations
can be found by solving the corresponding
determining {\it linear} operator equations.
In this sense, the symmetry properties of nearly linear equations are similar in many respects to those of
linear equations.

Let us consider brief\/ly a~method for solving the Cauchy problem~\eqref{cauchy-GPE} for the
RGPE~\eqref{QUAD-HAMILT}, following the scheme described in~\cite{shapovalov:BTS1}.
We denote the Weyl-ordered symbol of an operator $\hat A(t)=A(\hat z,t)$ by $A(z,t)$ and def\/ine the
expectation value for $\hat A(t)$ over the state $\Psi(\vec x,t)$ as
\begin{gather*}
A_\Psi(t)=\dfrac{1}{\|\Psi\|^2}(\Psi,\hat A(t)\Psi)=\dfrac{1}{\|\Psi\|^2}\int_{{\mathbb R}^n}
\D\vec x\Psi^*(\vec x,t)\hat A(t)\Psi(\vec x,t).
\end{gather*}

As $\|\Psi\|^2$ does not depend on time, we have from~\eqref{QUAD-HAMILT}, \eqref{shapovalov:QUAD-1},
and~\eqref{shapovalov:QUAD-2}
\begin{gather}
\dot{A}_\Psi(t)=\frac{1}{\|\Psi\|^2}\int_{{\mathbb R}^n}
\D\vec x\Psi^*(\vec x,t)\Bigg\{\frac{\partial\hat A(t)}{\partial t}+\frac{i}{\hbar}[{H_{\rm qu}(\hat z,t)},\hat A(t)]_{-}
\nonumber
\\
\phantom{\dot{A}_\Psi(t)=}
{}+\dfrac{i\tilde\varkappa}{\hbar}\int_{{\mathbb R}^n}\D\vec y\Psi^*(\vec y,t)[V_{\rm qu}
(\hat z,\hat w,t),\hat A(t)]_{-}\Psi(\vec x,t)\Bigg\},
\label{phaz1lst2_1}
\end{gather}
where $\dot{ A}_\Psi(t)= \D A_\Psi(t)/\D t$ and $\tilde\varkappa=\varkappa\|\Psi\|^2=\varkappa\|\psi\|^2$.

We call~\eqref{phaz1lst2_1} {\em the Ehrenfest equation} for the RGPE~\eqref{QUAD-HAMILT} as is common
practice in quantum mechanics for the linear Schr\"odinger equation ($\varkappa=0$
in~\eqref{shapovalov:GPE}).

Let $z_\Psi(t)=(z_{\Psi l}(t))$ and $\Delta^{(2)}_\Psi(t)=\big( \Delta^{(2)}_{\Psi  kl}(t)
\big)$ denote the expectation values over $\Psi(\vec x,t)$ for the operators
\begin{gather*}
\hat z_l,
\qquad
\hat{\Delta}^{(2)}_{kl}=\dfrac{1}{2}\big(\Delta\hat z_{k}\Delta\hat z_l+\Delta\hat z_l\Delta\hat z_k\big),
\qquad
k,l=\overline{1,2n},
\end{gather*}
respectively.
Here $\Delta\hat z_l =\hat z_l- (z_\Psi)_{l}(t)$.
We call $z_\Psi(t)$ the f\/irst moments and $\Delta^{(2)}_\Psi(t)$
the second centered moments of
$\Psi(\vec x,t)$.

From~\eqref{QUAD-HAMILT},~\eqref{shapovalov:QUAD-1},~\eqref{shapovalov:QUAD-2}, and~\eqref{phaz1lst2_1} we
immediately obtain a~dynamical system in matrix notation:
\begin{gather}
\dot z_{\Psi}=J\big\{{\mathcal H}_z(t)+[{\mathcal H}_{zz}(t)+\tilde\varkappa(W_{zz}(t)+W_{zw}(t))]z_{\Psi}\big\},
\nonumber
\\[8pt]
\dot\Delta_{\Psi}^{(2)}=J[{\mathcal H}_{zz}(t)+\tilde\varkappa W_{zz}(t)]\Delta_{\Psi}^{(2)}-\Delta_{\Psi}
^{(2)}[{\mathcal H}_{zz}(t)+\tilde\varkappa W_{zz}(t)]J.
\label{shapovalov:HES}
\end{gather}
We call~\eqref{shapovalov:HES} {\em the Hamilton--Ehrenfest system} (HES) of the second order for the
RGPE~\eqref{QUAD-HAMILT}
as~\eqref{shapovalov:HES} contain the f\/irst and second moments.

For brevity, we use a~shorthand notation for the total set of the f\/irst and second moments of $\Psi(\vec
x,t)$:
\begin{gather}
\label{phaz1lst2_2a_2m}
{\mathfrak g}_\Psi(t)=\big(z_{\Psi}(t),\Delta^{(2)}_{\Psi}(t)\big).
\end{gather}

The functions ${\mathfrak g}={\mathfrak g}_\Psi(t)$ describe phase orbits in the phase space of
system~\eqref{shapovalov:HES}.

Then the Cauchy problem~\eqref{cauchy-GPE} for the RGPE~\eqref{QUAD-HAMILT} can be written equivalently as
\begin{gather}
\hat L(t,{\mathfrak g}_\Psi(t))\Psi(\vec x,t)
=\big\{{-}i\hbar\partial_t+\hat H_{q}(t,{\mathfrak g}_\Psi(t))\big\}\Psi(\vec x,t)=0,
\label{shapovalov:GPE0}
\\
\hat H_{q}(t,{\mathfrak g}_\Psi(t))
=\dfrac{1}{2}\langle\hat z,{\mathcal H}_{zz}(t)\hat z\rangle
+\langle{\mathcal H}_z(t),\hat z\rangle
+\dfrac{\tilde\varkappa}{2}\langle\hat z,W_{zz}(t)\hat z\rangle
\nonumber
\\
\qquad
{}+\dfrac{\tilde\varkappa}{2}\langle z_{\Psi}(t),W_{ww}(t)z_{\Psi}(t)\rangle
+\tilde\varkappa\langle\hat z,W_{zw}(t)z_{\Psi}(t)\rangle
+\dfrac{\tilde\varkappa}{2}\Sp\Big[W_{ww}(t)\Delta_{\Psi}^{(2)}(t)\Big],
\label{shapovalov:GPE3}
\\
\dot{\mathfrak g}_\Psi(t)=\Gamma(t,{\mathfrak g}_\Psi(t)),
\label{phaz1lst2_2a}
\\
\Psi(\vec x,t)\Big|_{t=s}=\psi(\vec x),
\qquad
{\mathfrak g}_\Psi(t)\Big|_{t=s}={\mathfrak g}_\psi.
\label{shapovalov:GPE1}
\end{gather}
Equation~\eqref{phaz1lst2_2a} is a~concise form of the HES~\eqref{shapovalov:HES},
and $\Gamma(t,{\mathfrak g}_\Psi(t))$ designates the r.h.s.\ of~\eqref{shapovalov:HES}.

We call the reduced GPE~\eqref{shapovalov:GPE0} and the corresponding HES~\eqref{phaz1lst2_2a} {\it the
consistent system} for the RGPE~\eqref{QUAD-HAMILT}.
The reduced GPE~\eqref{shapovalov:GPE0} can be assigned to the class of nearly linear
equations~\cite{LevShTr12}, as the operator~\eqref{shapovalov:GPE3} of the RGPE~\eqref{shapovalov:GPE0} is
a~linear partial dif\/ferential operator with coef\/f\/icients depending only on the f\/irst and second
moments ${\mathfrak g}_\Psi(t)$.

The consistent system~\eqref{shapovalov:GPE0}, \eqref{phaz1lst2_2a} allows us to reduce the Cauchy problem
for the RGPE \eqref{shapovalov:GPE0} to the Cauchy problem for a~linear PDE,
therefore
the Cauchy problem~\eqref{shapovalov:GPE1} for HES~\eqref{phaz1lst2_2a} can be solved independently of equation~\eqref{shapovalov:GPE0}.

Let
\begin{gather*}
{\mathfrak g}(t,{\bf C})=\big(z(t,{\bf C}),\Delta^{(2)}(t,{\bf C})\big)
\end{gather*}
be the general solution of the HES~\eqref{phaz1lst2_2a} and ${\bf C}=(C_1,C_2,\dots, $ $C_N)$ denote the
set of integration constants.

Consider a~linear PDE with coef\/f\/icients depending on the parameters ${\bf C}$:
\begin{gather}
\hat L(t,{\bf C})\Phi(\vec x,t,{\bf C})=\big\{{-}i\hbar\partial_t+\hat H_{q}(t,{\bf C})\big\}
\Phi(\vec x,t,{\bf C})=0,
\label{shapovalov:GPE4}
\end{gather}
where
\begin{gather}
\hat H_{q}(t,{\bf C})=\dfrac{1}{2}\langle\hat z,{\mathcal H}_{zz}(t)\hat z\rangle+\langle{\mathcal H}
_z(t),\hat z\rangle+\dfrac{\tilde\varkappa}{2}\langle\hat z,W_{zz}(t)\hat z\rangle
+\tilde\varkappa\langle\hat z,W_{zw}(t)Z(t,{\bf C})\rangle
\nonumber
\\
\phantom{\hat H_{q}(t,{\bf C})=}
{}+\dfrac{\tilde\varkappa}{2}\langle Z(t,{\bf C}
),W_{ww}(t)Z(t,{\bf C})\rangle+\dfrac{\tilde\varkappa}{2}\Sp\Big[W_{ww}(t)\Delta^{(2)}(t,{\bf C})\Big].
\label{shapovalov:GPE5}
\end{gather}
The operator $\hat H_{q}(t,{\bf C})$
of~\eqref{shapovalov:GPE4} is obtained
from~\eqref{shapovalov:GPE3} where the general solution ${\mathfrak g}(t,{\bf C})$ of the
HES~\eqref{phaz1lst2_2a} stands for the moments ${\mathfrak g}_\Psi(t)$.
We call~\eqref{shapovalov:GPE4} {\it the associated linear equation} (ALE) for the
RGPE~\eqref{shapovalov:GPE0}.

Let $\Phi (\vec x,t,{\bf C}[\psi])$ denote the solution of the Cauchy problem for the
ALE~\eqref{shapovalov:GPE4} with the initial condition
\begin{gather}
\Phi(\vec x,t,{\bf C}[\psi])\Big{|}_{t=s}=\psi(\vec x),
\label{shapovalov:ALE-1}
\end{gather}
where the integration constants ${\bf C}$ have been replaced by the functionals ${\bf C}={\bf C}[\psi]$
determined from the algebraic conditions
\begin{gather}
{\mathfrak g}(t,{\bf C})\Big|_{t=s}={\mathfrak g}_\psi.
\label{shapovalov:GPE7}
\end{gather}

Then the solution of the Cauchy problem~\eqref{shapovalov:GPE0},~\eqref{shapovalov:GPE3} for the RGPE
(see~\cite{shapovalov:BTS1,Lisok:Sigma} for details)~is
\begin{gather}
\Psi(\vec x,t)=\Phi(\vec x,t,{\bf C}[\psi]).
\label{shapovalov:GPE6}
\end{gather}

Def\/ine ${\bf C}[\Psi](t)$ by the algebraic condition
\begin{gather}
{\mathfrak g}(t,{\bf C}[\Psi](t))={\mathfrak g}_\Psi(t).
\label{shapovalov:GPE7-1}
\end{gather}

From the uniqueness of the solution of the Cauchy problem for the HES~\eqref{shapovalov:GPE1} it follows
that
\begin{gather*}
{\mathfrak g}(t,{\bf C}[\Psi](t))={\mathfrak g}(t,{\bf C}[\psi])
\end{gather*}
and, hence,
\begin{gather}
{\bf C}[\Psi](t)={\bf C}[\psi],
\label{shapovalov:GPE8-1aa}
\end{gather}
i.e., the functionals ${\bf C}[\Psi](t)$ are the integrals of~\eqref{shapovalov:GPE}.

Also, we have
\begin{gather}
{\mathfrak g}(t,{\bf C}[\psi])={\mathfrak g}_\psi(t),
\label{shapovalov:GPE8-1a}
\end{gather}
where ${\mathfrak g}_\psi(t)$ is the solution of the HES~\eqref{phaz1lst2_2a} with the initial
condition~\eqref{shapovalov:GPE7}.

The 1D case of equation~\eqref{shapovalov:GPE8-1aa} is considered in more detail
in~\cite{shapovalov:LTS}.
Solving the associated linear equation \eqref{shapovalov:GPE4} with the algebraic condition~\eqref{shapovalov:GPE7}
we obtain a~solution to the nonlinear equation~\eqref{QUAD-HAMILT}.

Let us now turn to the construction of symmetry operators for the RGPE~\eqref{QUAD-HAMILT}.
By using an operator intertwining a~pair of ALEs of the form~\eqref{shapovalov:GPE4}.
Analysis of the GPE of general form involves a~great number of additional technical issues associated with
the semiclassical approximation that requires a~separate study.
To illustrate the main ideas of the proposed approach, we restrict our discussion to the case of
a~quadratic operator for which equation~\eqref{shapovalov:GPE4} is integrable.

\section{The intertwining operator and symmetry operators}\label{section3}

In this section, we establish a~relationship between the symmetry operators and the intertwining operator
for the reduced Gross--Pitavevskii equation~\eqref{QUAD-HAMILT}.
A class of intertwining operators can be found as a~set of products of the fundamental intertwining
operator by
the symmetry operators of the ALE~\eqref{shapovalov:GPE4}.

According to def\/inition~\eqref{symm-oper-1}, the nonlinear symmetry operator $\hat A(t)$ maps any
solution $\Psi (\vec x,t)$ of equation~\eqref{shapovalov:GPE0} into its another solution:
\begin{gather*}
\Psi_A(\vec x,t)=(\hat A(t)\Psi)(\vec x,t).
\end{gather*}
For $\hat a=\hat A(t)\big |_{t=s}$ and $\psi(\vec x)$ given by~\eqref{cauchy-GPE}, we can set
\begin{gather*}
\psi_a(\vec x)=\hat a\psi(\vec x)=\Psi_A(t)\big|_{t=s}
\end{gather*}
and use the notation ${\mathfrak g}_{\psi_a}$ for the f\/irst and second moments of $\psi_a(\vec x)$
similar to~\eqref{phaz1lst2_2a_2m}.

From the solution of the Cauchy problem for the HES~\eqref{shapovalov:GPE1} with the initial condition
${\mathfrak g}_\Psi(t) \Big|_{t=s}={\mathfrak g}_{\psi_a}$,
analogously to
\eqref{shapovalov:GPE8-1a}, we have
\begin{gather}
{\mathfrak g}(t,{\bf C}[\psi_a])={\mathfrak g}_{\psi_a}(t).
\label{shapovalov:GPE8-1ab}
\end{gather}

According to~\eqref{shapovalov:GPE6}, the solutions $\Psi(\vec x,t)$ and $\Psi_A(\vec x,t)$ of the
RGPE~\eqref{shapovalov:GPE0} are found as
\begin{gather}
\Psi(\vec x,t)=\Phi(\vec x,t,{\bf C})\big|_{{\bf C}={\bf C}[\psi]}
\label{psi-1}
\end{gather}
and
\begin{gather*}
\Psi_A(\vec x,t)=\Phi(\vec x,t,{\bf C'})\big|_{{\bf C'}={\bf C'}[\psi_a]},
\end{gather*}
where $\Phi(\vec x,t, {\bf C})$ and $\Phi(\vec x,t, {\bf C'})$ are the solutions of two ALEs of the
form~\eqref{shapovalov:GPE4} with two dif\/ferent sets of integration constants ${\bf C}$ and ${\bf C'}$,
respectively, and the corresponding linear operators $\hat L(t,\bf{C'})$ and $\hat L(t,{\bf C})$.

To construct the symmetry operator $\hat A(t)$
we relate the functions $\Phi(\vec x,t, {\bf C'}[\psi_a])$
and\linebreak
$\Phi (\vec x,t, {\bf C}[\psi])$ by a~linear operator $\hat M(t,s, {\bf C'},{\bf C})$ intertwining the
operators $\hat L(t,\bf{C'})$ and $\hat L(t,{\bf C})$:
\begin{gather}
\hat L(t,{\bf C}')\hat M(t,s,{\bf C}',{\bf C})=\hat R(t,s,{\bf C}',{\bf C})\hat L(t,{\bf C}).
\label{opred-eq-112}
\end{gather}
Here the linear operator $\hat R(t,s,{\bf C}',{\bf C})$ is a~Lagrangian multiplier, and the initial
condition is $\hat M(t,s,{\bf C}',{\bf C})|_{t=s}=\hat a$.

From~\eqref{opred-eq-112} we have that $\Phi(\vec x,t,{\bf C'})=\hat M(t,s,{\bf C'}, {\bf C})\Phi (\vec
x,t, {\bf C})$ for two arbitrary sets of constants ${\bf C'}$ and ${\bf C}$, and this is especially true
for $\Phi (\vec x,t, {\bf C'}[\psi_a])$ and $\Phi (\vec x,t, {\bf C}[\psi])$ with the constants ${\bf
C'}[\psi_a]$ and ${\bf C}[\psi]$.

To f\/ind the operator $\hat M(t,s, {\bf C'}, {\bf C})$, we consider a~linear intertwining operator
$\widehat{\mathcal D}(t,s,{\bf C}',{\bf C})$ for $\hat L(t,{\bf C}')$ and $\hat L(t,{\bf C})$ satisfying
the conditions
\begin{gather}
\hat L(t,{\bf C}')\hat{\mathcal D}(t,s,{\bf C}',{\bf C})=\hat{\mathcal D}(t,s,{\bf C}',{\bf C}
)\hat L(t,{\bf C}),
\label{splet1}
\\
\widehat{\mathcal D}(t,s,{\bf C}',{\bf C})\Big|_{t=s}=\hat{\mathbb I}.
\end{gather}

We call $\widehat{\mathcal D}(t,s,{\bf C}',{\bf C})$ {\it the fundamental intertwining operator} for $\hat
L(t,{\bf C}')$ and $\hat L(t,{\bf C})$.
Making use
of $\widehat{\mathcal D}(t,s,{\bf C}',{\bf C})$, the operator $\hat M(t,s,{\bf C}',{\bf C})$
involved
into~\eqref{opred-eq-112} can be presented as
\begin{gather*}
\hat M(t,s,{\bf C}',{\bf C})=\widehat{\mathcal D}(t,s,{\bf C}',{\bf C})\widehat{B}(t,{\bf C}).
\end{gather*}
Here $\widehat{B}(t,{\bf C})\in \mathcal{B}$ is the linear symmetry operator of
ALE~\eqref{shapovalov:GPE4} satisfying the conditions
\begin{gather}
[\hat L(t,{\bf C}),\widehat{B}(t,{\bf C})]_-=0,
\qquad
\widehat{B}(t,{\bf C})|_{t=s}=\hat a,
\label{opredm1}
\end{gather}
and $\mathcal{B}$ is the family of linear symmetry operators of the ALE~\eqref{opredm1}.

Hence, given the operator $\widehat{\mathcal D}(t,s,{\bf C}',{\bf C})$ of~\eqref{splet1} and the family
$\mathcal{B}$ of linear symmetry operators of the ALE~\eqref{opredm1} we can construct the family of
nonlinear symmetry operators for the GPE~\eqref{shapovalov:GPE}.

Thus, we arrive at
\begin{theorem}
Let $\hat{\mathcal D}(t,s,{\bf C}',{\bf C})$ be the fundamental intertwining operator~\eqref{splet1} for
$\hat L(t,{\bf C}')$ and $\hat L(t,{\bf C})$ and let $\widehat{B}(t,{\bf C})$ be the linear symmetry
operator of the ALE~\eqref{shapovalov:GPE4} satisfying the conditions~\eqref{opredm1}.
Then
\begin{gather}
(\hat A(t)\Psi)(\vec x,t)=\widehat{\mathcal D}\big(t,s,{\bf C'}[\hat a\psi],{\bf C}[\Psi](t)\big)
\widehat{B}(t,{\bf C}[\Psi](t))\Psi(\vec x,t)
\label{splet1r7}
\end{gather}
defines the family of nonlinear symmetry operators for the GPE~\eqref{QUAD-HAMILT}.
Here ${\bf C'}[\hat a\psi]$ and ${\bf C}[\Psi]$ $(={\bf C}[\psi])$ can be found
from~\eqref{shapovalov:GPE8-1a} and~\eqref{shapovalov:GPE7-1}, respectively, and $\widehat{B}\in
\mathcal{B}$.
\end{theorem}

Note that the symmetry operator $\hat A(t)$ from~\eqref{splet1r7} is nonlinear, as the operators
$\hat{\mathcal D}$ and $\hat B$ depend on the parameters $\bf C$ being functionals of the function $\Psi$.
To f\/ind the fundamental intertwining operator $\widehat{\mathcal D}(t,s,{\bf C}',{\bf C})$, we introduce
a~function $\phi(\vec x,t,{\bf C})$ by the conditions
\begin{gather}
\Phi(\vec x,t,{\bf C})=\hat K(\vec x,t,s,{\bf C})\phi(\vec x,t,{\bf C}),
\nonumber\\
\hat K(\vec x,t,s,{\bf C})=\exp[{-}\langle\vec X(t,{\bf C}),\nabla\rangle]\exp\Big\{\dfrac{i}
\hbar[S(t,{\bf C})+\langle\vec P(t,{\bf C}),\vec x\rangle]\Big\},
\label{Zamena}
\end{gather}
where $\Phi(\vec x,t,{\bf C})$ is a~solution of equation~\eqref{shapovalov:GPE4}, the vector $z=Z(t,{\bf C})=(\vec P(t,{\bf C}),\vec X(t,{\bf C}))$ satisf\/ies equation~\eqref{phaz1lst2_2a}, and $S(t,{\bf C})$ is a~smooth function to be determined.

For $\phi(\vec x,t,{\bf C})$ we have from~\eqref{shapovalov:GPE4}
\begin{gather*}
\hat L_0(\vec x,t,{\bf C})\phi(\vec x,t,{\bf C})=0,
\\
\hat L_0(\vec x,t,{\bf C})=\hat K^{-1}(\vec x,t,s,{\bf C})\hat L(\vec x,t,{\bf C}
)\hat K(\vec x,t,s,{\bf C})
\\
\phantom{\hat L_0(\vec x,t,{\bf C})}
=-i\hbar\partial_t+\langle\dot{\vec X}(t,{\bf C}
),i\hbar\nabla\rangle+\dot S(t,{\bf C})+\langle\dot{\vec{P}}(t,{\bf C}
),\vec x\rangle-\langle\vec P(t,{\bf C}),\dot{\vec{X}}(t,{\bf C})\rangle
\\
\phantom{\hat L_0(\vec x,t,{\bf C})=}
+\dfrac{1}{2}\langle(\hat z+Z(t,{\bf C})),{\mathcal H}_{zz}(t)(\hat z+Z(t,{\bf C}
))\rangle+\langle{\mathcal H}_z(t),(\hat z+Z(t,{\bf C}))\rangle
\\
\phantom{\hat L_0(\vec x,t,{\bf C})=}
+\tilde\varkappa\Big[\dfrac{1}{2}\langle(\hat z+Z(t,{\bf C})),W_{zz}(t)(\hat z+Z(t,{\bf C}
))\rangle+\langle(\hat z+Z(t,{\bf C})),W_{zw}(t)Z(t,{\bf C})\rangle
\\
\phantom{\hat L_0(\vec x,t,{\bf C})=}
+\dfrac{1}{2}\langle Z(t,{\bf C}),W_{ww}(t)Z(t,{\bf C})\rangle+\dfrac{1}{2}\Sp\big[W_{ww}(t)\Delta^{(2)}
(t,{\bf C})\big]\Big].
\end{gather*}
Putting
\begin{gather}
S(t,{\bf C})=\int_s^t\Bigl\lbrace\langle\vec P(t,{\bf C}),\dot{\vec X}(t,{\bf C})\rangle-{H}
_\varkappa(t,{\bf C})\Bigr\rbrace \D t,
\label{Zamena3}
\end{gather}
where
\begin{gather*}
{H}_\varkappa(t,{\bf C})=\dfrac12\langle Z(t,{\bf C}),[{\mathcal H}_{zz}(t)+\tilde\varkappa(W_{zz}
(t)+2W_{zw}(t)+W_{ww}(t))]Z(t,{\bf C})\rangle
\\
\phantom{{H}_\varkappa(t,{\bf C})=}
+\langle{\mathcal H}_z(t),Z(t,{\bf C})\rangle+\dfrac12\tilde\varkappa\text{Sp}[W_{ww}(t)\Delta^{(2)}
(t,{\bf C})],
\end{gather*}
and taking into account~\eqref{phaz1lst2_2a}, we obtain an equation for the function $\phi(\vec x,t,{\bf
C})$:
\begin{gather}
\hat L_0(\vec x,t)\phi(\vec x,t,{\bf C})=0,
\qquad
\hat L_0(\vec x,t)=-i\hbar\partial_t+\dfrac{1}{2}\langle\hat z,{\mathcal H}_{zz}
(t)\hat z\rangle+\tilde\varkappa\dfrac{1}{2}\langle\hat z,W_{zz}(t)\hat z\rangle.
\label{LinUravnenie}
\end{gather}

\begin{theorem}
Let
\begin{gather}
\hat b(t,s,{\bf C}',{\bf C})=\langle b(t,s,{\bf C}',{\bf C}),J\hat z\rangle
\label{splet1r6}
\end{gather}
and let the $2n$-component vector $b=b(t,s,{\bf C}',{\bf C})$ be a~solution of the Cauchy problem for the
system
\begin{gather*}
\dot b=J[{\mathcal H}_{zz}(t)+\tilde\varkappa W_{zz}(t)]b,
\qquad
b\big|_{t=s}=\delta Z_0({\bf C},{\bf C'}),
\\
\delta Z_0({\bf C},{\bf C'})=\big(\delta\vec P_0({\bf C},{\bf C'}),\delta\vec X_0({\bf C},{\bf C'}
)\big),
\\
\delta\vec X_0({\bf C},{\bf C'})=\vec X_0({\bf C'})-\vec X_0({\bf C}),
\qquad
\delta\vec P_0({\bf C},{\bf C'})=\vec P_0({\bf C'})-\vec P_0({\bf C}).
\end{gather*}
Then the operator $\widehat{\mathcal D} (t,s,{\bf C}',{\bf C})$ involved
into~\eqref{splet1} can be presented as
\begin{gather}
\widehat{\mathcal D}(t,s,{\bf C}',{\bf C})
=\exp\Big\{\frac{i}{2\hbar}\langle\delta\vec X_0({\bf C},{\bf C'}),\delta\vec P_0({\bf C},{\bf C'})\rangle\Big\}
\nonumber
\\
\phantom{\widehat{\mathcal D}(t,s,{\bf C}',{\bf C})=}
\times\hat K(\vec x,t,s,{\bf C'})\exp\Big\{\frac{i}{\hbar}\hat b(t,s,{\bf C}',{\bf C})\Big\}\hat K^{-1}
(\vec x,t,s,{\bf C}).
\label{splet1r1}
\end{gather}
The operator $\hat K(\vec x,t,s,{\bf C})$ is defined in~\eqref{Zamena}.
\end{theorem}

\begin{proof}
In view of~\eqref{Zamena} and~\eqref{LinUravnenie}, equation~\eqref{splet1} for the
fundamental intertwining operator can be written as
\begin{gather*}
\hat K(\vec x,t,s,{\bf C'})\hat L_0(\vec x,t)\hat K^{-1}(\vec x,t,s,{\bf C'})\hat{\mathcal D}(t,s,{\bf C}',{\bf C})
\\
\qquad
=\hat{\mathcal D}(t,s,{\bf C}',{\bf C})\hat K(\vec x,t,s,{\bf C})\hat L_0(\vec x,t)\hat K^{-1}
(\vec x,t,s,{\bf C}),
\qquad
\widehat{\mathcal D}(t,s,{\bf C}',{\bf C})\big|_{t=s}=\hat{\mathbb I}.
\end{gather*}
Therefore, the operator $\hat L_0(\vec x,t)$ given by~\eqref{LinUravnenie} and the function $\phi(\vec
x,t,{\bf C})$ do not depend on the constants $\bf C$.
Hence, we have
\begin{gather*}
\widehat{\mathcal D}(t,s,{\bf C}',{\bf C})
=\hat K(\vec x,t,s,{\bf C'})\widehat{\widetilde{\mathcal D}}
(t,s,{\bf C}',{\bf C})\hat K^{-1}(\vec x,t,s,{\bf C}),
\end{gather*}
where $\widehat{\widetilde{\mathcal D}}(t,s,{\bf C}',{\bf C})$ is the symmetry operator of
equation~\eqref{LinUravnenie}, i.e.\
\begin{gather}
\Big[\hat L_0(\vec x,t),\widehat{\widetilde{\mathcal D}}(t,s,{\bf C}',{\bf C})\Big]_-=0,\qquad
\widehat{\widetilde{\mathcal D}}(t,s,{\bf C}',{\bf C})\Big|_{t=s}
=\widehat{\widetilde{\mathcal D}}_0({\bf C}',{\bf C}).\label{splet1r2}
\end{gather}
Here we
used the notation
\begin{gather*}
\widehat{\widetilde{\mathcal D}}_0({\bf C}',{\bf C})
=\hat K^{-1}(\vec x,t,s,{\bf C'})\hat K(\vec x,t,s,{\bf C})\Big|_{t=s}
\\
\phantom{\widehat{\widetilde{\mathcal D}}_0({\bf C}',{\bf C})}
=\exp\Big[\langle\delta\vec X_0({\bf C},{\bf C'}),\nabla\rangle\!-\!\frac{i}\hbar\langle\delta\vec P_0({\bf C}
,{\bf C'}),\vec x\rangle\Big]\exp\Big\{\frac{i}{2\hbar}\langle\delta\vec X_0({\bf C},{\bf C'}
),\delta\vec P_0({\bf C},{\bf C'})\rangle\Big\}.
\end{gather*}
The solution of the Cauchy problem~\eqref{splet1r2} for the operator $\widehat{\widetilde{\mathcal
D}}(t,s,{\bf C}',{\bf C})$ can be obtained with the
standard methods (see, e.g.,~\cite{Bagre,shapovalov:MANKO})
as
\begin{gather*}
\widehat{\widetilde{\mathcal D}}(t,s,{\bf C}',{\bf C})=\exp\Big\{\frac{i}{2\hbar}
\langle\delta\vec X_0({\bf C},{\bf C'}),\delta\vec P_0({\bf C}',{\bf C})\rangle\Big\}\exp\Big\{\frac{i}
{\hbar}\hat b(t,s,{\bf C}',{\bf C})\Big\},\tag*{\qed}
\end{gather*}
\renewcommand{\qed}{}
\end{proof}

Then the symmetry operator $\hat A(t)$ for equation~\eqref{shapovalov:GPE0} (or, equivalently, for
equation~\eqref{QUAD-HAMILT}) can be presented as~\eqref{splet1r7}, where the intertwining operator
$\widehat{\mathcal D}(t,s,{\bf C}',{\bf C})$ is def\/ined by~\eqref{splet1r1} and $\widehat{B}(t,{\bf C})$
is the symmetry operator for the ALE~\eqref{shapovalov:GPE4}.

Using the explicit form~\eqref{splet1r1} of the intertwining operator $\widehat{\mathcal D}(t,s,{\bf
C}',{\bf C})$ and the operator $\hat K(\vec x,t,s,{\bf C})$ from~\eqref{Zamena}, we have
\begin{gather}
\Psi_A(\vec x,t)=(\hat A(t)\Psi)(\vec x,t)
\nonumber
\\
\phantom{\Psi_A(\vec x,t)}
=\exp\Big\{\dfrac{i}
\hbar[S_A(t)+\langle\vec P_A(t),\vec x-\vec X_A(t)\rangle]\Big\}\widehat{B}
\big(\vec x+\vec X(t)-\vec X_A(t),t\big)
\nonumber
\\
\phantom{\Psi_A(\vec x,t)=}
\times\exp\Big\{{-}\dfrac{i}\hbar[S(t)+\langle\vec P(t),\vec x-\vec X(t)\rangle]\Big\}
\Psi\big(\vec x+\vec X(t)-\vec X_A(t),t\big),
\label{lisok:eq01}
\end{gather}
where
\begin{gather*}
\widehat{B}(\vec x,t)=\widehat{B}\big(t,{\bf C}[\Psi](t)\big).
\end{gather*}

Note that expression~\eqref{lisok:eq01} for the symmetry operators is not simple and requires further
analysis, but other forms of symmetry operators for GPEs are unknown.

To obtain simplier examples of symmetry operators in explicit form, we consider the 1D case of
equations~\eqref{QUAD-HAMILT},~\eqref{shapovalov:QUAD-1}, and~\eqref{shapovalov:QUAD-2}.

\section{Symmetry operators in the 1D case}\label{section4}

Based on the results of the previous section, here we construct
in explicit form
the symmetry operators for the RGPE~\eqref{QUAD-HAMILT} in the one-dimensional case
and obtain two countable sets of exact
solutions to the one-dimensional GPE using
the symmetry operators.

Consider the reduced 1D GPE~\eqref{QUAD-HAMILT}
\begin{gather}
\hat F(\Psi)(\vec x,t)=\big\{{-}i\hbar\partial_t+\widehat H_{\rm qu}+\varkappa\widehat V_{\rm qu}
(\Psi)(t)\big\}\Psi(x,t)=0,
\label{spgau2no}
\\
\Psi\big|_{t=0}=\psi(x),
\label{spgau2noa}
\end{gather}
where we
used the notations
\begin{gather*}
\widehat{H}_{\rm qu}=\frac12\left(\mu\hat p^2+\rho(x\hat p+\hat px)+\sigma x^2\right),
\qquad
\widehat V_{\rm qu}(\Psi)=\frac12\int_{-\infty}^{+\infty}\D y\left(ax^2+2bxy+cy^2\right)|\Psi({y}
)|^2,
\end{gather*}
{}$\hat p=-i\hbar\partial/\partial x$; $a$, $b$, and $c$ are the real parameters of the nonlocal operator
$\widehat V_{\rm qu}(\Psi)$; $\mu$, $\sigma$, and $\rho$ are the parameters of the linear operator
$\widehat{H}_{\rm qu}$; $x, y\in \mathbb{R}^1$.

The Hamilton--Ehrenfest system~\eqref{phaz1lst2_2a} for the f\/irst-order moments becomes~\cite{BLT07}
\begin{gather}
\dot{p}=-\rho p-\sigma_0x,
\nonumber
\\
\dot{x}=\mu p+\rho x,
\label{spgau2n}
\end{gather}
and for the second-order moments with $\Delta_{21}^{(2)}=\Delta_{12}^{(2)}$ we have
\begin{gather}
\dot{\Delta}_{11}^{(2)}=-2\rho\Delta_{11}^{(2)}-2\tilde{\sigma}\Delta_{21}^{(2)},
\nonumber\\
\dot{\Delta}_{21}^{(2)}=\mu\Delta_{11}^{(2)}-\tilde{\sigma}\Delta_{22}^{(2)},
\nonumber\\
\dot{\Delta}_{22}^{(2)}=2\mu\Delta_{21}^{(2)}+2\rho\Delta_{22}^{(2)},
\label{spgau3n}
\end{gather}
where
\begin{gather*}
\sigma_{0}=\sigma+\tilde\varkappa(a+b),
\qquad
\tilde{\sigma}=\sigma+\tilde\varkappa a.
\end{gather*}
We introduce the notation
\begin{gather*}
\bar\Omega=\sqrt{\sigma_0\mu-\rho^2},
\qquad
\Omega=\sqrt{\tilde\sigma\mu-\rho^2}
\end{gather*}
and assume that $\bar\Omega^2=\sigma_0\mu-\rho^2>0$.
Indeed, in this case, the general solution of system~\eqref{spgau2n} is
\begin{gather}
X(t,{\bf C})=C_1\sin\bar\Omega t+C_2\cos\bar\Omega t,
\nonumber\\
P(t,{\bf C})=\frac{1}{\mu}\big(\bar\Omega C_1-\rho C_2\big)\cos\bar\Omega t-\frac{1}{\mu}
\big(\bar\Omega C_2+\rho C_1\big)\sin\bar\Omega t,
\label{gau15a}
\end{gather}
and all solutions of system~\eqref{spgau2n} are localized.

Assume that the wave packets that describe the evolution of particles by equation~\eqref{spgau2no} do not spread.
This takes place if $ \Omega^2 = \tilde\sigma\mu-\rho^2> 0 $.

For system~\eqref{spgau3n} we have
\begin{gather}
\Delta_{22}^{(2)}(t,{\bf C})=C_3\sin2\Omega t+C_4\cos2\Omega t+C_5,
\nonumber
\\
\Delta_{21}^{(2)}(t,{\bf C})=\frac{1}{\mu}(\Omega C_3-\rho C_4)\cos2\Omega t-\frac{1}{\mu}
(\Omega C_4+\rho C_3)\sin2\Omega t-\frac{\rho}{\mu}C_5,
\nonumber
\\
\Delta_{11}^{(2)}(t,{\bf C})=\frac{1}{\mu^2}\big(\big(\rho^2-\Omega^2\big)C_3+2\rho\Omega C_4\big)\sin2\Omega t
\nonumber
\\
\phantom{\Delta_{11}^{(2)}(t,{\bf C})=}
{}+\frac{1}{\mu^2}\big(\big(\rho^2-\Omega^2\big)C_4-2\rho\Omega C_3\big)\cos2\Omega t+\frac{\tilde\sigma}{\mu}C_5,
\label{gau15d}
\end{gather}
and all solutions of system~\eqref{spgau3n} are also localized.
Here ${\bf C}=(C_1,\ldots,C_5)$ and $C_l$, $l=\overline{1,5}$, are arbitrary integration constants.

The 1D associated
linear equation~\eqref{shapovalov:GPE4} is
\begin{gather}
\hat L(t,{\bf C})\Phi(\vec x,t,{\bf C})=\big\{{-}i\hbar\partial_t+\hat H_{q}(t,{\bf C})\big\}
\Phi(\vec x,t,{\bf C})=0,
\label{LinUravnenie_p1}
\\
\hat H_{q}(t,{\bf C})=\dfrac{\mu\hat p^2}{2}+\frac{\tilde{\sigma}x^2}{2}+\frac{\rho(x\hat p+\hat px)}
2+\tilde\varkappa bx X(t,{\bf C})+\tilde\varkappa\dfrac{c}{2}\Big[X^2(t,{\bf C})+\Delta_{22}^{(2)}(t,{\bf C})\Big].
\nonumber
\end{gather}

We can immediately verify that for the associated linear equation~\eqref{LinUravnenie_p1} we can construct
the following set of symmetry operators linear in $x$ and $\hat p$:
\begin{gather}
\hat a(t,{\bf C})
=\dfrac{1}{\sqrt{2\hbar}}\big[C(t)\big(\hat p-P(t,{\bf C})\big)-B(t)\big(x-X(t,{\bf C}
)\big)\big],
\label{lambda1}
\\
\hat a^{+}(t,{\bf C})
=\dfrac{1}{\sqrt{2\hbar}}\big[C^*(t)\big(\hat p-P(t,{\bf C}
)\big)-B^*(t)\big(x-X(t,{\bf C})\big)\big].\label{lambda2}
\end{gather}

Here the functions $B(t)$ and $C(t)$ are solutions of the linear Hamiltonian system
\begin{gather}
\dot{B}=-\rho B-\tilde\sigma C,
\nonumber\\
\dot{C}=\mu B+\rho C.
\label{spgau2v}
\end{gather}
The Cauchy matrix ${\mathcal X}(t)$ for system~\eqref{spgau2v} can easily be found as
\begin{gather}
{\mathcal X}(t)=
\begin{pmatrix}
\cos\Omega t-\dfrac{\rho}{\Omega}\sin\Omega t & -\dfrac{1}{\mu\Omega}\big(\Omega^2+\rho^2\big)\sin\Omega t
\\
\dfrac\mu\Omega\sin\Omega t & \cos\Omega t+\dfrac{\rho}{\Omega}\sin\Omega t
\end{pmatrix},
\qquad
{\mathcal X}(t)\Big|_{t=0}={\mathbb I}_{2\times2}.\label{spgau2w}
\end{gather}

The set of solutions normalized by the condition~\cite{shapovalov:MASLOV-1}
\begin{gather}
B(t)C^*(t)-C(t)B^*(t)=2i
\label{spgau2w1}
\end{gather}
can be written as
\begin{gather}
B(t)={e^{i\Omega t}}\frac{(-\rho+i\Omega)}{\sqrt{\Omega\mu}},
\qquad
C(t)={e^{i\Omega t}}\sqrt{\frac{\mu}{\Omega}}.
\label{spgau2w2}
\end{gather}
Equation~\eqref{spgau2w1} results in the following commutation relations for the symmetry
operators~\eqref{lambda1} and~\eqref{lambda2}:
\begin{gather*}
\big[\hat a(t,{\bf C}),\hat a^{+}(t,{\bf C})\big]_-=1.
\end{gather*}

For the function $\phi$ given by~\eqref{Zamena} in the 1D case, we obtain
\begin{gather}
\Phi(x,t,{\bf C})=\hat K(\vec x,t,s,{\bf C})\phi(\vec x,t),
\\
\hat K(x,t,{\bf C})=\exp[{-}X(t,{\bf C})\partial_x]\exp\Big\{\frac{i}\hbar[S(t,{\bf C})+P(t,{\bf C})x]\Big\},
\label{Zamena_pp}
\end{gather}
where, according to~\eqref{Zamena3},
\begin{gather}
S(t,{\bf C})=\int_0^t\big\{P(t,{\bf C})\dot{X}(t)-{H}_\varkappa(t,{\bf C})\big\}\D t,
\label{Zamena3_p}
\\
{H}_\varkappa(t,{\bf C})=\dfrac{\mu}{2}P^2(t,{\bf C})+\frac{1}{2}X^2(t,{\bf C})
\big[\sigma_0+\tilde\varkappa(b+c)\big]+\rho P(t,{\bf C})X(t,{\bf C})+\tilde\varkappa\dfrac{c}{2}
\Delta_{22}^{(2)}(t,{\bf C}).
\nonumber
\end{gather}
From~\eqref{LinUravnenie} we f\/ind
\begin{gather*}
\hat L_0(x,t)\phi=0,
\qquad
\hat L_0(x,t)=-i\hbar\partial_t+\dfrac{\mu\hat p^2}{2}+\dfrac{(\sigma+\tilde\varkappa a)x^2}{2}
+\frac{\rho(x\hat p+\hat px)}2.
\end{gather*}

Then the symmetry operator $\hat A(t)$~\eqref{splet1r7} for equation~\eqref{spgau2no} can be presented as
\begin{gather}
(\hat A(t)\Psi)(x,t)=\widehat{\mathcal D}(t,{\bf C}[\hat a\psi],{\bf C}[\Psi](t))\widehat{B}(t,{\bf C}
[\Psi](t))\Psi(x,t),
\label{splet1r7p}
\end{gather}
where $\widehat{B}(t,{\bf C})$ is the symmetry operator of the associated linear
equation~\eqref{LinUravnenie_p1}.

The intertwining operator $\widehat{\mathcal D}(t,{\bf C}',{\bf C})$ presented, according
to~\eqref{splet1r1}, as
\begin{gather}
\widehat{\mathcal D}(t,{\bf C}',{\bf C})=\exp\left\{i\frac{C_2'-C_2}{2\hbar\mu}
\Big(\bar\Omega(C_1'-C_1)-\rho(C_2'-C_2)\Big)\right\}
\nonumber
\\
\phantom{\widehat{\mathcal D}(t,{\bf C},{\bf C}')=}
\times\hat K^{-1}(x,t,{\bf C'})\exp\Big\{\frac{i}{\hbar}\hat b(t,{\bf C}',{\bf C})\Big\}\hat K(x,t,{\bf C}),
\label{splet1r1p}
\end{gather}
where, according to~\eqref{splet1r6},
\begin{gather*}
\hat b(t,{\bf C}',{\bf C})=b_x(t,{\bf C}',{\bf C})\hat p-b_p(t,{\bf C}',{\bf C})x
=\langle b(t),J\hat z\rangle,
\end{gather*}
and the vector $b(t)$ is def\/ined by
\begin{gather*}
b(t,{\bf C}',{\bf C})=
\begin{pmatrix}
b_p(t,{\bf C}',{\bf C})
\\
b_x(t,{\bf C}',{\bf C})
\end{pmatrix}
=\frac1\mu{\mathcal X}(t)
\begin{pmatrix}
\bar\Omega(C_1'-C_1)-\rho(C_2'-C_2)
\\
\mu(C_2'-C_2)
\end{pmatrix}
.
\end{gather*}
The matrix ${\mathcal X}(t)$ is given by~\eqref{spgau2w}.

The symmetry operator $\hat A (t)$ of the nonlinear equation~\eqref{spgau2no} involved
into~\eqref{splet1r7p}
has the structure of a~linear pseudodif\/ferential operator whose parameters are
functionals of the function on which the operator acts.
Therefore, the explicit form of the operator $\hat A(t)$ is determined not only
by the symmetry operator $\widehat{B}(t, {\bf C})$ of the associated linear equation, but also by the function~$\Psi(x, t)$.
Note that for some values of the parameters (more precisely, for the function~$\Psi(x, t)$ that def\/ines
them) the pseudodif\/ferential operator becomes a~dif\/ferential one.

We set{\samepage
\begin{gather}
\widehat{B}(t,{\bf C})=\widehat{B}_\nu(t,{\bf C})=\frac{1}{\sqrt{{\nu}!}}\big[\hat a^{+}(t,{\bf C})\big]^\nu,
\qquad
\nu\in{\mathbb Z}_+,
\label{gau15dpp}
\end{gather}
where the operator $\hat a^{+}(t,{\bf C})$
is def\/ined in~\eqref{lambda2}.}

Substituting~\eqref{gau15dpp} in~\eqref{splet1r7p}
we obtain the symmetry operator, which we denote by $\hat A_\nu(t)$.
Using
a~stationary solution of the Hamilton--Ehrenfest system~\eqref{spgau2n},~\eqref{spgau3n}
we simplify the symmetry operators~\eqref{lambda1},~\eqref{lambda2} and generate a~countable set of
explicit solutions of the 1D GPE~\eqref{spgau2no}.

A stationary solution of equations \eqref{spgau2n},~\eqref{spgau3n}
is obtained from the general solution~\eqref{gau15a},~\eqref{gau15d} if we
take integration constants as ${\bf C}={\bf C}^0=(C_1^0,\ldots,C_5^0)$,
where $C_1^0=C_2^0=C_3^0=C_4^0=0$ and~$C_5^0$ is an arbitrary real constant.
The stationary solution is
\begin{gather}
X(t,{\bf C})=P(t,{\bf C})=0,
\qquad
\Delta_{22}^{(2)}(t,{\bf C})=C_5^0,
\nonumber
\\
\Delta_{21}^{(2)}(t,{\bf C})=-\frac{\rho}{\mu}C_5^0,
\qquad
\Delta_{11}^{(2)}(t,{\bf C})=\frac{\tilde\sigma}{\mu}C_5^0.
\label{gau15dp}
\end{gather}

Substituting~\eqref{gau15dp} in~\eqref{LinUravnenie_p1}, we obtain the associated linear equation
\begin{gather}
\hat L\big(x,t,{\bf C}^0\big)\Phi=0,
\nonumber
\\
\hat L(x,t,{\bf C}^0)=\left[{-}i\hbar\partial_t+\dfrac{\mu\hat p^2}{2}+\dfrac{(\sigma+\tilde\varkappa a)x^2}{2}
+\frac{\rho(x\hat p+\hat px)}2+\tilde\varkappa\dfrac{c}{2}C_5^0\right].
\label{LinUravnenie_p2}
\end{gather}
The operator $\hat K(\vec x,t,{\bf C})=\hat K(\vec x,t,{\bf C}^0)$ from~\eqref{Zamena_pp} is the operator
of multiplication by the function
\begin{gather*}
\hat K(\vec x,t,{\bf C}^0)=\exp\Big\{{-}\frac{i}{2\hbar}\tilde\varkappa c C_5^0t\Big\}.
\end{gather*}

The linear operators~\eqref{lambda1} and~\eqref{lambda2} then become
\begin{gather}
\label{lambda}
\hat a(t,{\bf C}^0)=\frac{1}{\sqrt{2\hbar}}\big[C(t)\hat p-B(t)x\big],
\qquad
\hat a^{+}(t,{\bf C}^0)=\frac{1}{\sqrt{2\hbar}}\big[C^*(t)\hat p-B^*(t)x\big];
\end{gather}
they are symmetry operators for equation~\eqref{LinUravnenie_p2}; the functions $C(t)$ and $B(t)$ are def\/ined
in~\eqref{spgau2w2}.

The function
\begin{gather}
\Phi_{0}(x,t,{\bf C}^0)=\left(\frac{1}{\pi{\hbar}}\right)^{1/4}\left(\frac{\Omega}{\mu}\right)^{1/4}
\exp\left\{{-}\frac{i}{2{\hbar}}\frac{\rho}{\mu}x^2-\frac{1}{2{\hbar}}\frac{\Omega}{\mu}x^2\right\}
\nonumber
\\
\phantom{\Phi_{0}(x,t,{\bf C}^0)=}
\times
\exp\left\{{-}\frac{i}{2}\Omega t-\frac{i}{2\hbar}\tilde\varkappa c C_5^0t\right\}
\label{lambda3}
\end{gather}
is easily verif\/ied to be a~solution of equation~\eqref{LinUravnenie_p2}.

Upon direct substitution, we see that for the function~\eqref{lambda3},
equations~\eqref{shapovalov:ALE-1},~\eqref{shapovalov:GPE7}, which determine the functionals ${\bf
C}[\Psi](t)$, become
\begin{gather}
X(0,{\bf C})=x_\psi=0,
\qquad
P(0,{\bf C})=p_\psi=0,
\nonumber
\\
\Delta_{22}^{(2)}(0,{\bf C})=\big(\Delta_{22}^{(2)}\big)_\psi=\frac{\hbar}{2}\big|C(0)\big|^2=\frac{\hbar\mu}
{2\Omega},
\nonumber
\\
\Delta_{11}^{(2)}(t,{\bf C})=\big(\Delta_{11}^{(2)}\big)_\psi=\frac{\hbar}{2}
\big|B(0)\big|^2=\frac{\hbar(\varrho^2+\Omega^2)}{2\Omega\mu},
\label{a21_q}
\\
\Delta_{12}^{(2)}(0,{\bf C})=\big(\Delta_{12}^{(2)}\big)_\psi=\frac{\hbar}{4}
\big[B(0)C^*(0)+B^*(0)C(0)\big]=-\frac{\hbar\varrho}{2\Omega},
\nonumber
\\[1.2mm]
\psi(x)=\Phi_{0}\big(x,0,{\bf C}^0\big).
\nonumber
\end{gather}

From~\eqref{a21_q} and~\eqref{gau15dp} it follows that ${ C}^0_5=(\hbar\mu/2{\Omega})$.
From~\eqref{psi-1} and~\eqref{lambda3} we f\/ind a~particular solution $\Psi_{0}(x,t)$ of the
GPE~\eqref{spgau2no}:
\begin{gather}
\Psi_{0}(x,t)=\Phi_{0}\big(x,t,{\bf C}^0\big)\bigg|_{{C}^0_5=(\hbar\mu/2{\Omega})}
\nonumber
\\
\phantom{\Psi_{0}(x,t)}
=\left(\frac{1}{\pi{\hbar}}\right)^{1/4}\left(\frac{\Omega}{\mu}\right)^{1/4}
\exp\left\{{-}\frac{i}{2{\hbar}}\frac{\rho}{\mu}x^2-\frac{1}{2{\hbar}}\frac{\Omega}{\mu}x^2\right\}
\exp\left\{{-}\frac{i}{2}\Omega t-\frac{i\mu}{4{\Omega}}\tilde\varkappa c t\right\}.\label{lambda4}
\end{gather}

The symmetry operators~\eqref{gau15dpp},~\eqref{lambda} generate from
\eqref{lambda3} the solutions of the associated
linear equation~\eqref{LinUravnenie_p2} that constitute a~Fock basis in the space $L_2({\mathbb R})$:
\begin{gather}
\Phi_{\nu}\big(x,t,{\bf C}^0{}'\big)=\widehat{B}_\nu\big(t,{\bf C^0}'\big)\Phi_{0}\big(x,t,{\bf C}^0{}'\big)
=\frac{1}{\sqrt{{\nu}!}}\Big[\hat a^{+}\big(t,{\bf C}^0{}'\big)\Big]^\nu\Phi_{0}\big(x,t,{\bf C}^0{}'\big)
\nonumber
\\
\phantom{\Phi_{\nu}\big(x,t,{\bf C}^0{}'\big)}
=\frac{i^{\nu}}{\sqrt{{\nu}!}}\left(\frac{1}{\sqrt{2}}\right)^{\nu}H_{\nu}
\biggl(\sqrt{\frac{\Omega}{{\hbar}\mu}}x\biggr)\Phi_{0}\big(x,t,{\bf C}^0{}'\big)\exp\big\{{-}i\Omega\nu t\big\},
\qquad
\nu\in{\mathbb Z}_+,\label{a21_q2}
\end{gather}
where $H_{\nu}(\zeta)$ are the Hermite polynomials~\cite{Beitman2}
\begin{gather*}
H_{\nu}(\zeta)=\bigg(2\zeta-\dfrac{\D}{\D\zeta}\bigg)^{\nu}\cdot1.
\end{gather*}

The operator~\eqref{splet1r1p}, intertwining the operators $\hat L( x,t,{\bf C}^0)$ and $\hat L( x,t,{\bf
C}^0{}')$ of the form~\eqref{LinUravnenie_p2}, reads
\begin{gather*}
\widehat{\mathcal D}\big(t,{\bf C}^0{}',{\bf C}^0\big)=\exp\left\{\frac{i}{2\hbar}\tilde\varkappa c\big[C_5^0-C_5^0{}'\big]t\right\}.
\end{gather*}
Equations~\eqref{shapovalov:GPE7} that determine the functionals ${\bf C}[\Psi_\nu](t)$ for the
functions~\eqref{a21_q2} can be written~as
\begin{gather}
X(t,{\bf C})=x_{\psi_a}=0,
\qquad
P(0,{\bf C})=p_{\psi_a}=0,
\nonumber
\\
\Delta_{22}^{(2)}(0,{\bf C})=\big(\Delta_{22}^{(2)}\big)_{\psi_a}=\dfrac{\hbar}{2}
(2\nu+1)\big|C(0)\big|^2=\dfrac{\hbar\mu}{2\Omega}(2\nu+1),
\nonumber
\\
\Delta_{11}^{(2)}(0,{\bf C})=\big(\Delta_{11}^{(2)}\big)_{\psi_a}=\dfrac{\hbar}{2}
\big|B(0)\big|^2(2\nu+1)=\dfrac{\hbar(\rho^2+\Omega^2)}{2\Omega\mu}(2\nu+1),\label{a21_p}
\\
\Delta_{12}^{(2)}(0,{\bf C})=\big(\Delta_{12}^{(2)}\big)_{\psi_a}=\dfrac{\hbar}{4}
\big[B(0)C^*(0)+B^*(0)C(0)\big](2\nu+1)=-\dfrac{\hbar\rho}{2\Omega}(2\nu+1),
\nonumber
\\
{\psi_a}(x)=\Phi_{\nu}\big(x,0,{\bf C}^0{}'\big)=\widehat{B}_\nu\big(0,{\bf C^0}'\big)\psi(x).
\nonumber
\end{gather}

Here we have used the standard properties of Hermite polynomials~\cite{Beitman2}.
Taking into account~\eqref{gau15dp}, we f\/ind from~\eqref{a21_p} that ${
C}_5^0{}'=(\hbar\mu/{\Omega})({\nu}+{1}/{2})$.

Then the symmetry operator $\hat A_\nu(t)$ determined by~\eqref{splet1r7p} transforms the solution
$\Psi_{0}(x,t)$ of~\eqref{lambda4} into a~solution $\Psi_{\nu}(x,t)$ of the nonlinear GPE~\eqref{spgau2no}
according to the following relation:
\begin{gather}
\Psi_{\nu}(x,t)=(\hat A_\nu(t)\Psi_0)(x,t)
\nonumber
\\
\phantom{\Psi_{\nu}(x,t)}
=\widehat{\mathcal D}\big(t,{\bf C}^0,{\bf C}^0{}'\big)\dfrac{1}{\sqrt{{\nu}!}}\Big[\hat a^{+}(t,{\bf C}
^0)\Big]^\nu\bigg|_{{C}^0_5=(\hbar\mu/2{\Omega}),{C}_5^0{}'=(\hbar\mu/{\Omega})({\nu}+{1}/{2})}\Psi_{0}
(x,t)
\nonumber
\\
\phantom{\Psi_{\nu}(x,t)}
=\left(\dfrac{1}{\sqrt{2}}\right)^{\nu}\left(\dfrac{1}{\pi{\hbar}}\right)^{1/4}\left(\frac{\Omega}{\mu}
\right)^{1/4}\exp\Big\{{-}\dfrac{i}{2{\hbar}}\frac{\rho}{\mu}x^2-\dfrac{1}{2{\hbar}}\dfrac{\Omega}{\mu}
x^2\Big\}
\nonumber
\\
\phantom{\Psi_{\nu}(x,t)=}
\times H_{\nu}\biggl(\sqrt{\dfrac{\Omega}{{\hbar}\mu}}x\biggr)\exp\Big\{{-}i\Big({\nu}+\dfrac{1}{2}
\Big)\Big(\dfrac{\tilde\varkappa c\mu}{2\Omega}+\Omega\Big)t\Big\}.
\label{spsob1}
\end{gather}

The functions~\eqref{spsob1} constitute a~countable set of particular solutions to
equation~\eqref{spgau2no} which are generated from $\Psi_{0}(x,t)$ by the nonlinear symmetry operator $\hat A_\nu(t)$.

The symmetry operators $\hat A_\nu(t)$
in~\eqref{spsob1} generalize those
of the linear equations used in the Maslov complex germ theory~\cite{shapovalov:BEL-DOB, shapovalov:MASLOV-1}, as in the
limit $\varkappa \to 0$ ($\varkappa $ is the nonlinearity parameter in equation~\eqref{spgau2noa}),
the operators $\hat A_\nu(t)$ become the creation operators of the Maslov complex germ theory.
As in the linear case ($\varkappa = 0$), the operators $\hat A_\nu(t)$ generate a~countable set of exact
solutions $\Psi_{\nu}(x,t)$ to the nonlinear equation~\eqref{spgau2noa}.

Assume that ${\bf C}={\bf C}^1=(C_1^1,C_2^1,0,0,C_5^1)$.
This choice of the constants yields the following expression for the phase
orbit~\eqref{gau15a},~\eqref{gau15d}:
\begin{gather}
X\big(t,C_1^1,C_2^1\big)=C_1^1\sin\bar\Omega t+C_2^1\cos\bar\Omega t,
\nonumber
\\
P\big(t,C_1^1,C_2^1\big)=\dfrac{1}{\mu}\big(\bar\Omega C_1^1-\rho C_2^1\big)\cos\bar\Omega t-\frac{1}{\mu}
\big(\bar\Omega C_2^1+\rho C_1^1\big)\sin\bar\Omega t,
\label{gau15dppp}
\\
\Delta_{22}^{(2)}\big(t,{\bf C}^1\big)=C_5^1,
\qquad
\Delta_{21}^{(2)}\big(t,{\bf C}^1\big)=-\dfrac{\rho}{\mu}C_5^1,
\qquad
\Delta_{11}^{(2)}\big(t,{\bf C}^1\big)=\dfrac{\tilde\sigma}{\mu}C_5^1.
\nonumber
\end{gather}

In view of~\eqref{gau15dppp} and~\eqref{Zamena_pp}, we have
\begin{gather*}
\hat K(x,t,{\bf C}^1)
=\exp\big[{-}X\big(t,{\bf C}^1\big)\partial_x\big]\exp\Big\{\frac{i}\hbar\big[S\big(t,{\bf C}^1\big)+P\big(t,{\bf C}^1\big)x\big]\Big\}.
\end{gather*}

Consider the action of the operator $\hat A_0 (t)$ involved
into~\eqref{splet1r7p},~\eqref{gau15dpp} on the
functions~\eqref{spsob1}.
Let us write the operator~\eqref{splet1r1p}
intertwining the operators $L( x,t,{\bf C}^0)$ and $L(x,t,{\bf C}^1)$ determined by~\eqref{LinUravnenie_p2}
as
\begin{gather*}
\widehat{\mathcal D}\big(t,{\bf C},{\bf C}^0\big)
=\exp[{-}X(t,{\bf C})\partial_x]\exp\left\{\frac{i}\hbar\left[S(t,{\bf C})
-\frac{1}{2}\tilde\varkappa c C_5^0t+P\big(t,{\bf C}^1\big)x\right]\right\}
\\
\phantom{\widehat{\mathcal D}\big(t,{\bf C},{\bf C}^0\big)=}
\times\exp\Big\{\frac{i}{2\hbar\mu}C_2\big(\tilde{\Omega}C_1-\rho C_2\big)\Big\}\exp\Big\{\frac{i}{\hbar}
\hat b\big(t,{\bf C}^1,{\bf C}^0\big)\Big\},
\end{gather*}
where
\begin{gather*}
\hat b\big(t,{\bf C},{\bf C}^0\big)=b_x\big(t,{\bf C},{\bf C}^0\big)\hat p-b_p(t,{\bf C},{\bf C}
^0)x=\langle b(t),J\hat z\rangle,
\end{gather*}
and the vector $b(t)$ is def\/ined by the expression
\begin{gather*}
b\big(t,{\bf C},{\bf C}^0\big)=
\begin{pmatrix}
b_p\big(t,{\bf C},{\bf C}^0\big)
\\
b_x\big(t,{\bf C},{\bf C}^0\big)
\end{pmatrix}
=\frac1\mu{\mathcal X}(t)
\begin{pmatrix}
\tilde{\Omega}C_1-\rho C_2
\\
\mu C_2
\end{pmatrix}
.
\end{gather*}
The matrix ${\mathcal X}(t)$ is given by~\eqref{spgau2w}.

Let us construct a~nonlinear symmetry operator
$\hat A(t,\alpha)$ corresponding to the nonstationary phase
orbit~\eqref{gau15a},~\eqref{gau15d}.
The operator $\hat A(t,\alpha)$ maps the nonstationary solution of equation~\eqref{spgau2no},
$\Psi_{\nu}(x,t)$ given by~\eqref{spsob1}, into another nonstationary solution of this equation,
$\widetilde\Psi_{\nu}(x,t)$.
Consider the shift operator
\begin{gather}
\widehat{B}\big(t,{\bf C}^1\big)=\widehat{B}\big(t,\alpha,{\bf C}^1\big)=\exp\{\alpha\hat{a}{}^+(t)-\alpha^*\hat{a}
(t)\},
\qquad
\alpha\in{\mathbb C},
\label{gau15dpppp}
\end{gather}
where the operators $\hat a(t,{\bf C})$ and $\hat a^{+}(t,{\bf C})$ are def\/ined by
expressions~\eqref{lambda1},~\eqref{lambda2}.
The operator~\eqref{gau15dpppp} should be substituted in~\eqref{splet1r7p} for the symmetry operator
$\widehat{B}(t,{\bf C})$.

Let us write the operator $\widehat{B}(t,\alpha,{\bf C}^1)$ involved
into~\eqref{gau15dpppp} as
\begin{gather*}
\widehat{B}(t,\alpha,{\bf C}^1)=\exp\{\beta(t)\hat{p}+\gamma(t)x\}=\exp\Big\{{-}\frac{i\hbar}
2\beta(t)\gamma(t)\Big\}\exp\{\gamma(t)x\}\exp\{\beta(t)\hat{p}\},
\end{gather*}
where
\begin{gather*}
\beta(t)=\dfrac1{\sqrt{2\hbar}}\big[C^*(t)\alpha-C(t)\alpha^*\big],
\qquad
\gamma(t)=\dfrac{1}{\sqrt{2\hbar}}\big[B(t)\alpha^*-B^*(t)\alpha\big].
\end{gather*}
Thus, we have
\begin{gather}
{\psi_a}(x)=\widehat{B}\big(0,\alpha,{\bf C}^1\big)\Phi_{\nu}\big(x,0,{\bf C}^0{}'\big)
\nonumber
\\
\phantom{{\psi_a}(x)}
=\exp\Big\{{-}\dfrac{i\hbar}2\beta(0)\gamma(0)\Big\}\exp\{\gamma(0)x\}\Phi_{\nu}
\big(x-i\hbar\beta(0),0,{\bf C}^0{}'\big),
\label{lst03prf12h}
\\
\Phi_{\nu}\big(x-i\hbar\beta(0),0,{\bf C}^0{}'\big)
=\frac{i^{\nu}}{\sqrt{{\nu}!}}\left(\frac{1}{\sqrt{2}}\right)^{\nu}H_{\nu}
\biggl(\sqrt{\frac{\Omega}{{\hbar}\mu}}\Big[x-i\hbar\beta(0)\Big]\biggr)\Phi_{0}\big(x-i\hbar\beta(0),0,{\bf C}^0{}'\big),
\nonumber
\\
\Phi_{0}\big(x-i\hbar\beta(0),0,{\bf C}^0{}'\big)
\nonumber
\\
\phantom{{\psi_a}(x)}
=\Phi_{0}\big(x,0,{\bf C}^0{}'\big)\exp\left\{\left[\left(\frac{-\rho+i\Omega}{\mu}\right)x\beta(0)-i\dfrac{\hbar}
2\left(\frac{-\rho+i\Omega}{\mu}\right)\beta^2(0)\right]\right\}.
\nonumber
\end{gather}
Note that
\begin{gather*}
\gamma(0)=\dfrac{1}{\sqrt{2\hbar}}\big[{-}B^*(0)\alpha+B(0)\alpha^*\big]
=\dfrac{1}{\sqrt{2\hbar}}
\left[\dfrac{(\rho+i\Omega)}{\sqrt{\Omega\mu}}\alpha+\frac{(-\rho+i\Omega)}{\sqrt{\Omega\mu}}\alpha^*\right]
\\
\phantom{\gamma(0)}
=i\dfrac{\sqrt{2}}{\sqrt{\hbar\Omega\mu}}\big[\rho\alpha_2+\Omega\alpha_1\big],
\\
\beta(0)=\dfrac1{\sqrt{2\hbar}}\big[C^*(0)\alpha-C(0)\alpha^*\big]=\dfrac1{\sqrt{2\hbar}}\sqrt{\frac{\mu}
{\Omega}}(\alpha-\alpha^*)=i\sqrt{\dfrac{2\mu}{\hbar\Omega}}\alpha_2,
\\
\gamma(0)+\dfrac{-\rho+i\Omega}{\mu}\beta(0)=\dfrac1{\sqrt{2\hbar}}
\left[{-}B^*(0)\alpha+B(0)\alpha^*+\dfrac{-\rho+i\Omega}{\mu}\big(C^*(0)\alpha-C(0)\alpha^*\big)\right]
\\
\phantom{\gamma(0)}
=\dfrac1{\sqrt{2\hbar\Omega\mu}}
\big[(\rho+i\Omega)\alpha+(-\rho+i\Omega)\alpha^*\big)
+(-\rho+i\Omega)(\alpha-\alpha^*)\big]=\dfrac{2i\sqrt{\Omega}\alpha}{\sqrt{2\hbar\mu}}.
\end{gather*}
Here $\alpha_1=\Re\alpha$ and $\alpha_2=\Im\alpha$.
Similarly, we have
\begin{gather}
-\frac{i\hbar}2\left[\beta(0)\gamma(0)+\frac{-\rho+i\Omega}{\mu}\beta^2(0)\right]
=-\frac{i\hbar}2\dfrac{2i\sqrt{\Omega}\alpha}{\sqrt{2\hbar\mu}}\beta(0)
=\dfrac12\alpha\big[\alpha-\alpha^*\big]
\nonumber
\\
\hphantom{-\frac{i\hbar}2\left[\beta(0)\gamma(0)+\frac{-\rho+i\Omega}{\mu}\beta^2(0)\right]}{}
=\frac12\big[\alpha^2-|\alpha|^2\big]=i\alpha\alpha_2.
\label{lst03prf15h}
\end{gather}
Substituting~\eqref{lst03prf15h} in~\eqref{lst03prf12h}, we obtain
\begin{gather}
\psi_a(x)=\widehat{B}\big(0,\alpha,{\bf C}^1\big)\Phi_{\nu}(x,0,{\bf C}^1)
\nonumber
\\
\phantom{\psi_a(x)}
=\exp\left\{\dfrac{i\alpha\alpha_2}2\right\}\exp\left\{\dfrac{2i\sqrt{\Omega}}{\sqrt{2\hbar\mu}}\alpha x\right\}
\dfrac{i^{\nu}}{\sqrt{\nu!}}\left(\frac{1}{\sqrt{2}}\right)^{\nu}H_{\nu}\bigg(\sqrt{\dfrac{\Omega}{\hbar\mu}
}\Big[x+\dfrac{\sqrt{2\hbar\mu}}{\Omega}\alpha_2\Big]\bigg)
\nonumber
\\
\phantom{\psi_a(x)=}
\times\left(\frac{1}{\pi{\hbar}}\right)^{1/4}\left(\frac{\Omega}{\mu}\right)^{1/4}
\exp\left\{\frac{i}{2{\hbar}}\left(\frac{-\rho+i\Omega}{\mu}\right)x^2\right\}.
\label{lst03prf12hh}
\end{gather}

From~\eqref{lst03prf12hh}, in particular, it follows that
\begin{gather*}
|{\psi_a}(x)|^2=\sqrt{\frac{\Omega}{\pi\hbar\mu}}\exp\left\{{-}\frac{\Omega}{\hbar\mu}
\left(x+\frac{\sqrt{2\hbar\mu}}{\sqrt{\Omega}}\alpha_2\right)^2\right\}
\frac{1}{{\nu}!}\Big(\frac{1}{2}\Big)^{\nu}H_{\nu}^2\bigg(\sqrt{\frac{\Omega}{\hbar\mu}}
\left[x+\frac{\sqrt{2\hbar\mu}}{\sqrt{\Omega}}\alpha_2\right]\bigg).
\end{gather*}

Similar to~\eqref{a21_p}, we write equations~\eqref{shapovalov:GPE7-1} determining the functionals ${\bf
C}[\psi_a](t)$ for the functions~\eqref{lst03prf12hh} as
\begin{gather*}
X\big(0,C_1^1,C_2^1\big)=C_2^1=x_\psi=-\frac{\sqrt{2\hbar\mu}}{\sqrt{\Omega}}\alpha_2,
\\
P\big(0,C_1^1,C_2^1\big)=\frac{1}{\mu}\big(\bar\Omega C_1^1-\rho C_2^1\big)=p_\psi=\dfrac{\sqrt{2\hbar\Omega}}{\sqrt{\mu}}
\alpha_1,
\\
\Delta_{22}^{(2)}(0,{\bf C})=\big(\Delta_{22}^{(2)}\big)_\psi=\frac{\hbar}{2}
(2\nu+1)\big|C(0)\big|^2=\frac{\hbar\mu}{2\Omega}(2\nu+1),
\\
\Delta_{11}^{(2)}(0,{\bf C})=\big(\Delta_{11}^{(2)}\big)_\psi=\frac{\hbar}{2}
\big|B(0)\big|^2(2\nu+1)=\frac{\hbar(\varrho^2+\Omega^2)}{2\Omega\mu}(2\nu+1),
\\
\Delta_{12}^{(2)}(0,{\bf C})=\big(\Delta_{12}^{(2)}\big)_\psi=\frac{\hbar}{4}
\big[B(0)C^*(0)+B^*(0)C(0)\big](2\nu+1)=-\frac{\hbar\varrho}{2\Omega}(2\nu+1).
\end{gather*}

From equations~\eqref{gau15a} and~\eqref{spgau2w1}, in view of~\eqref{shapovalov:GPE8-1ab}, we obtain
\begin{gather}
{C}_1^1=C_1^1(\alpha)=\dfrac{\sqrt{2\hbar\mu\Omega}}{\bar\Omega}\Big(\alpha_1-\dfrac\rho\Omega\alpha_2\Big),
\qquad
{C}_2^1=C_2^1(\alpha)=-\frac{\sqrt{2\hbar\mu}}{\sqrt\Omega}\alpha_2,
\nonumber\\
{C}_3^1={C}_4^1=0,
\qquad
{C}_5^1=\dfrac{\hbar\mu}{2\Omega}(2{\nu}+{1}).
\label{a21_pgh}
\end{gather}

Then the nonlinear symmetry operator for the nonlinear GPE~\eqref{spgau2no}, $\hat A(t,\alpha)$
determined by~\eqref{splet1r7p} and~\eqref{gau15dpppp}
transforms the solution $\Psi_{\nu}(x,t)$~\eqref{spsob1} into
a~nonstationary solution $\widetilde\Psi_{\nu}(x,t)$:
\begin{gather}
\widetilde\Psi_{\nu}(x,t,\alpha)=(\hat A(t,\alpha)\Psi_\nu)(x,t)=\widehat{\widetilde{\mathcal D}}
\big(t,{\bf C}^1,{\bf C}^0\big)\bigg|_{C_5^0={C}_5^1=(\hbar\mu/{\Omega})({\nu}+{1}/{2}),}\Psi_{\nu}(x,t)
\nonumber
\\
\phantom{\widetilde\Psi_{\nu}(x,t,\alpha)}
=\frac{i^{\nu}}{\sqrt{{\nu}!}}\left(\frac{1}{\sqrt{2}}\right)^{\nu}\left(\frac{1}{\pi{\hbar}}\right)^{1/4}
\left(\frac{\Omega}{\mu}\right)^{1/4}\exp\left\{{-}\frac{i}{2{\hbar}}\frac{\rho}{\mu}\Delta x^2-\frac{1}{2{\hbar}}
\frac{\Omega}{\mu}\Delta x^2\right\}
\nonumber
\\
\phantom{\widetilde\Psi_{\nu}(x,t,\alpha)=}
\times H_{\nu}\biggl(\sqrt{\frac{\Omega}{{\hbar}\mu}}\Delta x\biggr)\exp\Big\{{-}i\Big({\nu}
+\frac{1}{2}\Big)\Big(\frac{\tilde\varkappa c\mu}{2\Omega}+\Omega\Big)t\Big\}
\nonumber
\\
\phantom{\widetilde\Psi_{\nu}(x,t,\alpha)=}
\times\exp\Big\{\frac{i}{\hbar}
\big[S(t,C_1^1(\alpha),C_2^1(\alpha))+P(t,C_1^1(\alpha),C_2^1(\alpha))\Delta x\big]\Big\},
\label{spsob1h}
\end{gather}
which is localized around the phase orbit $\big(P\big(t,C_1^1(\alpha),C_2^1(\alpha)\big)$,
$X\big(t,C_1^1(\alpha),C_2^1(\alpha)\big)\big)$.
Here $\Delta x=x-X(t,C_1^1(\alpha),C_2^1(\alpha))$ with the constants $C_1^1(\alpha)$, $C_2^1(\alpha)$
determined by equation~\eqref{a21_pgh}, and the function $S(t,C_1^1(\alpha),C_2^1(\alpha))$ is determined
by~\eqref{Zamena3_p} where $C_1=C_1^1 (\alpha)$, $C_2=C_2^1(\alpha)$, $C_3=0$, $C_4=0$, and
$C_5^0=(\hbar\mu/{\Omega})({\nu}+{1}/{2})$.

In the linear case ($\varkappa=0$), the operators~\eqref{spsob1h} with $\alpha\in\mathbb C$ form
a~representation of the Heisenberg--Weyl group~\cite{shapovalov:MANKO,Perel}.
The functions $\widetilde\Psi_{\nu}(x,t,\alpha)$ determined by~\eqref{spsob1h} minimize the Schr\"odinger
uncertainty relation for $\nu=0$~\cite{Robertson34}, and, hence, they describe squeezed coherent
states~\cite{DodKurManko86}.

\section{Discussion}

Direct calculation of symmetry operators for a nonlinear equation
is, as a~rule, a~severe problem
because of the nonlinearity and complexity of the determining equations~\cite{shapovalov:PUKHNACHEV}.
However, for nearly linear equations~\cite{LevShTr12}
a~wide class of symmetry operators can be
constructed by solving linear determining equations for operators of this type
much as
symmetry operators are found for linear PDEs.
We have illustrated this situation
with
the example of the generalized multidimensional Gross--Pitaevskii equation~\eqref{shapovalov:GPE}.
The formalism of semiclassical asymptotics leads to the semiclassically reduced GPE~\eqref{QUAD-HAMILT}
(or~\eqref{shapovalov:GPE0}), which belongs to the class of nearly linear equations.
Note that the solutions of GPE can be found in a~special class of functions decreasing at
inf\/inity~\cite{shapovalov:BTS1}.
The reduced GPE is the quadratic one
in the space coordinates and derivatives and contains a~nonlocal term
of special form.
In constructing the symmetry operators for the reduced Gross--Pitaevskii equation~\eqref{shapovalov:GPE0},
we use the fact that this equation can be associated with the linear equation~\eqref{shapovalov:GPE4}.
The symmetry operator $\hat A (t)$ of the reduced GPE~\eqref{shapovalov:GPE0}, which
is a particular case
of~\eqref{splet1r7}, has the structure of a~linear pseudo\-dif\-ferential operator with coef\/f\/icients
$\bf C$ depending on the function $\Psi$ on which the operator acts.
The operator $\hat A$ is determined in terms of the linear intertwining operator $\widehat{\mathcal{D}}$
and of the symmetry operators of the associated linear equation~\eqref{shapovalov:GPE4}.
The dependence
of the coef\/f\/icients $\bf C$ on $\Psi$ arises from the algebraic
condition~\eqref{shapovalov:GPE8-1ab}, and therefore the operator $\hat A (t) $ is nonlinear.
This is the key point
of the presented approach.
The 1D examples considered show that for a~special choice of the parameters $\bf C$
we can construct
symmetry operators and generate the families of solutions to the nonlinear equation~\eqref{spgau2no}
written in explicit form.

The further development of the study of symmetry operators is seen as a~generalization to the approach for
integro-dif\/ferential GPEs of more general form and to systems of equations of this type.

\subsection*{Acknowledgements}

We would like to thank the anonymous referees who gave a~relevant contribution to improve the paper.
The work was supported in part by the Russian Federation programs ``Kadry'' (contract No.~16.740.11.0469)
and ``Nauka'' (contract No.~1.604.2011) and by Tomsk State University project No.~2.3684.2011.

\pdfbookmark[1]{References}{ref}
\LastPageEnding

\end{document}